\documentclass[12pt]{article}
\pdfoutput=1

\usepackage{graphicx} 
\usepackage{geometry}               

\usepackage[parfill]{parskip}    
\usepackage{amssymb,amsmath,bm,bbm,amsthm}
\usepackage{amsfonts}
\usepackage{amsbsy}
\usepackage{epstopdf}
\usepackage{xcolor}
\usepackage{float}
\usepackage{xy,tikz}
\usepackage{nicematrix}
\usepackage{arydshln}
\usepackage{mathtools}
\usepackage{dutchcal}
\usepackage{wrapfig}

\newcommand{\half}{{\textstyle \frac12}}
\newcommand{\ihalf}{{\textstyle \frac{i}2}}

\newcommand{\pa}{\partial}
\newcommand{\na}{\nabla}
\newcommand{\de}{\delta}
\newcommand{\si}{$\sigma$}
\newcommand{\om}{\omega}

\newcommand{\al}{\alpha}
\newcommand{\be}{\beta}

\newcommand{\te}{\theta}

\newcommand{\la}{\lambda}
\newcommand{\La}{\Lambda}

\newcommand{\vf}{{z}}
\newcommand{\nsa}{{n_{\!A}}}
\newcommand{\bvf}{\cbar{z}}

\newcommand{\tk}{{\widetilde K}}
\newcommand{\tr}{\,{\mathrm {tr}}}

\newcommand{\beq}{\begin{equation}}
\newcommand{\eeq}{\end{equation}}
\newcommand{\bea}{\begin{eqnarray}}
\newcommand{\eea}{\end{eqnarray}}
\newcommand{\lra}{\leftrightarrow}
\newcommand{\then}{~~\Rightarrow~~}
\newcommand{\nn}{\non}
\newcommand{\mcl}[1]{{\mathcal{#1}}}
\newcommand{\LL}{{\mathscr{L}}}
\newcommand{\bmn}{{\mathbcal{D}}}

\newcommand{\bfx}{{\mathbf{x}}}
\newcommand{\bbg}{{\mathbb{G}}}
\newcommand{\bbr}{{\mathbb{R}}}
\newcommand{\bbd}{{\mathbb{D}}}
\newcommand{\bbc}{{\mathbb{C}}}
\newcommand{\bbz}{{\mathbb{Z}}}

\newcommand{\bbbd}{\cbar{\mathbb{D}}}
\newcommand{\cp}{{\mathbb{C} \mathrm{P}}}

\newcommand{\lab}{\label}
\newcommand{\cbar}[1]{\mkern 1.5mu\overline{\mkern-1.5mu#1\mkern-1.5mu}\mkern 1.5mu}

\newcommand{\non}{\nonumber\\}

\newcommand{\bbD}[1]{\mathbb{D}_{#1}}
\newcommand{\bbDB}[1]{\cbar{\mathbb{D}}_{#1}}

\NiceMatrixOptions
  {
    custom-line = 
     {
       letter = I , 
       command = hdashedline , 
       tikz = dashed ,
       width = \pgflinewidth
     }
  }

\usepackage{setspace}

\usepackage[bookmarks=true,colorlinks=true,linkcolor=blue,citecolor=blue,urlcolor=blue,bookmarksnumbered]{hyperref}

\usepackage[
    backend=bibtex,
    style=numeric,
    sorting=none,
    maxbibnames=99,
    giveninits=true,
    minalphanames=1,
    maxalphanames=3,url=false
  ]{biblatex}
   
  \bibliography{RicciToric}
  
\renewbibmacro{in:}{}

\usepackage{mathabx}

\newbibmacro{string+doi}[1]{%
  \iffieldundef{doi}{#1}{\href{http://dx.doi.org/\thefield{doi}}{#1}}}
\DeclareFieldFormat{title}{\usebibmacro{string+doi}{\mkbibemph{#1}}}
\DeclareFieldFormat[article]{title}{\usebibmacro{string+doi}{\mkbibquote{#1}}}

\numberwithin{equation}{section}

\usepackage{mathrsfs}
\allowdisplaybreaks
\title{
\vspace{-2.5cm}
  \hfill{\normalfont\small UUITP-20/26}\\[-6mm]
  \hfill{\normalfont\small YITP-SB-2026-14}\\[1.5cm]
\textbf{Ricci-flat metrics from gauged linear} \\ \textbf{\si-models without RG flow}}

\author{ \hspace{-1cm}
Dmitri Bykov$^{\,a,\,b,\,c,\,d}$\footnote{Email:
 bykov@mi-ras.ru, dmitri.v.bykov@gmail.com} \qquad  Ond\v rej Hul\' ik$^{\,e , \, i}$\footnote{Email: ondra.hulik@gmail.com} \qquad Ulf Lindstr\"om$^{\,f,\,g}$\footnote{Email:  ulf.lindstrom@physics.uu.se} \\ 
 \hspace{-1cm} Martin Ro\v cek$^{\,h}$\footnote{Email: martin.rocek@stonybrook.edu} \qquad Rikard von Unge$^{\,i}$\footnote{Email: unge@physics.muni.cz} 
\\  \vspace{0cm}  \\ 
\hspace{-1cm}\vspace{-2mm} {\footnotesize $a)$ 
\emph{Steklov
Mathematical Institute of Russian Academy of Sciences, Gubkina str. 8, 119333 Moscow, Russia}} \\ \hspace{-1cm}\vspace{-2mm}
{\footnotesize $b)$ 
\emph{Institute for Theoretical and Mathematical Physics, Lomonosov Moscow State University, 119991 Moscow, Russia}} 
\\ \hspace{-1cm}\vspace{-2mm}
{\footnotesize $c)$ \emph{HSE University, 6 Usacheva str., Moscow 119048, Russia}}\\
\hspace{-1cm}\vspace{-2mm}
{\footnotesize $d)$ \emph{Beijing Institute of Mathematical Sciences and Applications (BIMSA), Huairou District, Beijing
101408, China}} 
\\
\hspace{-1cm}\vspace{-2mm}
{\footnotesize $e)$ \emph{
Institute for Mathematics Ruprecht-Karls-Universit\"at Heidelberg, 69120 Heidelberg, Germany}}\\ 
\hspace{-1cm}\vspace{-2mm}
{\footnotesize $f)$ \emph{
Department of Physics and Astronomy, Division of Theoretical Physics,}}\\ {\footnotesize \vspace{-2mm}\emph{Uppsala University,
Box 516, SE-75120 Uppsala, Sweden}}\\
\hspace{-1cm}\vspace{-2mm}
{\footnotesize $g)$ \emph{Center for Geometry and Physics, Uppsala University, Box 480, SE-75106 Uppsala, Sweden}}\\
\hspace{-1cm}\vspace{-2mm}
{\footnotesize $h)$ \emph{C.N. Yang Institute for Theoretical Physics, Stony Brook University, Stony Brook NY 11794-3840, USA}}\\
\hspace{-1cm}\vspace{-2mm}
{\footnotesize $i)$ \emph{Department of Theoretical Physics and Astrophysics
Faculty of Science,}}\\ 
{\footnotesize \vspace{-2mm}\emph{Masaryk University
Kotl\'a\v{r}sk\'a 2, CS-61137 Brno, Czechia}}
}
\date{}

\begin{document}

\maketitle
~\\
\textbf{Abstract:}
We investigate a class of Ricci-flat Kähler metrics on generalized conifolds constructed via gauged linear sigma models (GLSMs) with indefinite signature. By introducing “shadow” coordinates (superfields) entering the sigma model with negative signature kinetic term, we show that these GLSMs yield explicit Ricci-flat metrics on complex cones over products of projective spaces. We provide a general formula for the resulting Kähler potentials, along with detailed examples. Our results suggest new directions for the study of Calabi–Yau metrics and toric geometry, and raise interesting questions about the geometric meaning of indefinite signature models. We also give an interpretation in terms of a novel generalized K\"ahler gauging.\\~\\

\eject
\tableofcontents

\section{Introduction}

Ricci-flat K\"ahler manifolds constitute a fundamental ingredient in many constructions of string theory. For example, they arise naturally in compactifications of ten-dimensional spacetimes that preserve a portion of the original supersymmetry~\cite{CandelasWitten}, as well as in the effective actions for D-branes probing orbifold singularities~\cite{Douglas:1997de,Douglas:1997zj}. Such manifolds are often referred to as Calabi-Yau, though in the present paper Calabi-Yau will refer, more generally, to K\"ahler manifolds with vanishing first Chern class. By the conjecture of Calabi~\cite{CalabiRicci} proved by Yau~\cite{Yau1, Yau2}, these admit a unique Ricci flat metric in every K\"ahler class. 

An important task is to construct explicit examples of Ricci-flat K\"ahler metrics; an incomplete list of results in this direction includes~\cite{EH, GH, CdO, Pedersen,  Stenzel, BiquardGauduchon, PandoZayasTseytlin, Nitta1, Nitta2, Nitta3, Nitta4, Gauduchon,Martelli:2007pv, Chen:2006xh, vanCoevering, AZadeBykov}. In the special case of hyperk\"ahler manifolds (which are also Ricci-flat) this can sometimes be done via the hyperk\"ahler quotient construction~\cite{HKLR}. In the general case one can use the much less restrictive K\"ahler quotient; in physics terminology K\"ahler quotients of flat space are described by gauged linear sigma models (GLSM's). However, even in cases when the manifold in question is Calabi-Yau, metrics constructed via the K\"ahler quotient are, in general, \textit{not} Ricci-flat. In those cases the non-zero $\beta$-function is believed to drive the metric to a Ricci-flat one via renormalization group (Ricci) flow in the IR limit~\cite{Witten, AharonyFlow, ChenShifman}.   In this article, we observe that metrics on various generalizations of conifolds constructed by gauged linear \si-models of 
spaces with {\em indefinite} signature can be tuned to give Ricci-flat metrics.\footnote{Finding nice descriptions of Riemannian metrics by taking a quotient of an indefinite signature metric is well known, e.g., for hyperbolic spaces. See the comment after~(\ref{Lcp1}) below.}

Although our setup is more general, here we will only consider GLSM's with Abelian gauge groups, i.e., we restrict to the realm of \textit{toric} geometry. In practice this will mean that there are as many $U(1)$ isometries as there are chiral superfields, so that the K\"ahler potential\footnote{Recall that the K\"ahler potential serves as the superspace Lagrangian for models  with $(2,2)$ supersymmetry.} is independent of the phases of these chiral fields. 
In this case the $\sigma$-model may be dualized into a description in terms of real linear superfields, which gives us a complementary way to discuss the geometry. The new Lagrangian is the Legendre transform of the original K\"ahler potential, called the symplectic potential~\cite{Guillemin, Abreu}, and it describes the geometry in ``action-angle" variables. The domain where the symplectic potential is defined  is the moment polytope of the toric manifold, so that the potential concisely encodes both differential-geometric and topological information about the manifold. As we shall explain, the symplectic potential also serves as the K\"ahler potential of the T-dual K\"ahler manifold. 

One way of constructing explicit Ricci-flat metrics is the Calabi ansatz that gives the K\"ahler potential for a Ricci-flat metric on a holomorphic line bundle over a K\"ahler-Einstein base~\cite{Calabi1979}. Generalizing the Calabi ansatz, we find a GLSM that  produces the Ricci-flat metric on the canonical line bundle over any product of complex projective spaces. Somewhat surprisingly it inevitably involves chiral fields with the wrong sign kinetic term. We call such superfields {\em shadow} fields; despite their role in the construction, the resulting geometries are described by conventional positive-definite metrics. An interesting reinterpretation arises when we replace {\em shadow} fields by twisted chiral fields and the K\"ahler quotient by a generalized K\"ahler quotient, thus entering the realm of generalized K\"ahler geometry. We explain that such a replacement allows working with positive-definite metrics at all stages.

The paper is organized as follows: In section \ref{susysigma} we review the gauged linear sigma model construction in $(2,2)$ superspace. First discussing sigma models in general and then concentrating on the appearance of the Ricci-flatness condition and its meaning, we also introduce the shadow fields and give a motivation for doing so. In section \ref{fans} we then introduce the toric geometry needed for the paper and generalize it to include the newly introduced shadow fields. The main result of the paper is given in section \ref{CalAn} where we use a generalization of Calabi's ansatz to derive Ricci-flat metrics on complex cones over products of projective spaces and interpret the resulting sigma models as GLSMs. Finally, in section \ref{ssshadow}, we give a physical
interpretation of the indefinite signature linear sigma models by introducing twisted chiral superfields and writing our starting point as a quotient of a generalized Kähler target space with a generalized isometry implemented using a Large Vector multiplet. We summarize our results and discuss future directions in the conclusion. Finally we collect some basic facts about $(2,2)$ superspace, as well as details of the solution of the Monge-Amp\`ere equation, in the Appendix.

\section{Superspace sigma models}\lab{susysigma}
We start by reviewing the superspace formulation of sigma models in general and gauged linear sigma models in particular.
Toric geometry arises for the special case of gauged linear \si-models where the gauge group is Abelian. 

\subsection{Gauged Linear sigma models}
We consider $D=2$ \si-models in $(2,2)$ superspace, which means that we describe geometries using K\"ahler potentials\footnote{A brief review of $(2,2)$ superspace and its relation to complex geometry is given in Appendix \ref{geosuprev}.}. A {\em linear} \si-model has a set of chiral superfields $x^i, ~i=1\ldots K$ that give complex coordinates on flat $\bbc^K$:
\beq\lab{free}
\LL=\sum_i^K|x^i|^2~.
\eeq
The superspace Lagrange density $L$ is just the K\"ahler
potential for the flat metric on $\bbc^K$. We will actually need a slight generalization, and consider models with an indefinite signature target space $\bbc^{K,M}$:
\beq\lab{frees}
\LL=\sum_{i=1}^K|x^i|^2-\sum_{i=K+1}^{K+M}|x^i|^2\equiv \sum_{i=1}^{K+M}\eta_i|x^i|^2~~,
\eeq
where $\eta_i$ is $+1$ for the {\em physical} fields (or coordinates) $x^i,~i\le K$ and $-1$ for the {\em shadow} fields (or coordinates) $x^i,~i>K$. The shadow fields might seem to cause trouble with unitarity but we always consider models where the resulting geometry of the quotient is positive definite.

\subsubsection{GLSM basics}
A Kähler quotient is the Kähler–geometry analogue of symplectic reduction; it is a way to construct new Kähler manifolds by quotienting a Kähler manifold by a symmetry group, while preserving the Kähler structure. To ensure this, the group action needs to be both Hamiltonian and holomorphic. A convenient way to achieve this is to rely on supersymmetry. A sigma model in $(2,2)$ superspace is manifestly supersymmetric and thereby guarantees that the target space is Kähler. Any operation performed on this model, if it does not break supersymmetry, is sure to preserve the Kähler structure of the target space~\cite{Zumino}. A Kähler quotient is realized by superymmetrically gauging a sigma model isometry and then integrating out the gauge field.

\subsubsection{The charge matrix {\em Q} and the K\"ahler quotient}
The GLSM's we consider in this paper are Abelian\footnote{We consider gauge symmetries that act linearly and holomorphically on $\bbc^{K}$.}: the quotient is by a torus action $U(1)^m$ acting on $\bbc^{K}$. The action of the torus is specified by a $m\times K$ charge matrix $Q$; if we choose a basis for the $U(1)^m$ then we can write the action on each chiral superfield $x^i$ as:
 \beq \lab{expquo}
x^i \mapsto \left(e^{i\sum\limits_a\te_a Q^a_{\;\;i}} \right)x^i ~~,
 \eeq
where $a\in\{1\dots m\}$ and $i\in\{1\dots K\}$. The charge matrix $Q^a_{\;\;i}$ depends on the choice of basis for the torus acting on $\bbc^{K}$ as well as the ordering of the $x^i$, which means different models are determined by equivalence classes of charge matrices under permutation of the columns and linear combinations of the rows, i.e., the action of ${\rm GL}(m,\bbz)\times S_{K}$.

In superspace, the gauging is performed by introducing a real superfield $V_a$ for each gauge symmetry; however, no kinetic terms for the gauge fields are added. The resulting superspace Lagrangian $\LL$ gives rise to a K\"ahler quotient when all the $V_a$ are eliminated by extremizing $\LL$ with respect to $V_a$. In addition, 
we specify levels $c^a$ of the moment maps for each symmetry; in superspace, these are given by FI-terms, additional gauge-invariant terms that can be added without making the gauge fields dynamical. Then the gauged version of (\ref{frees}) is:
\bea
\LL = \sum_i \eta_i|x^i|^2 e^{\sum\limits_aV_a Q^a_i} - \sum_a V_a c^a~~.
\eea

\noindent{\bf Example:}
As an illustrative example, let us construct the Kähler potential of $\cp^1$ through a Kähler quotient of a gauged linear sigma model. Our starting point will be the model
\beq
\LL = |x^1|^2 + |x^2|^2
\eeq
where the fields have charges $Q=(1\;1)$. Gauging by introducing the gauge superfield $V$ and introducing the Fayet-Illiopoulos parameter $c$ we get
\beq\lab{exCP1}
\LL = e^V(|x^1|^2 + |x^2|^2) - cV
\eeq
Varying with respect to $V$ gives the equation $e^V = \frac{c}{|x^1|^2 + |x^2|^2}$, which, up to an additive constant, when inserted back into the Lagrangian gives
\beq
\LL =  c \ln (|x^1|^2 + |x^2|^2)
\eeq
We may finally use the gauge invariance to put $x^1 = 1$ so the final theory is given by
\beq\label{Lcp1}
\LL = c\ln ( 1 + |z|^2)
\eeq
where we recognize the Kähler potential for the Fubini-Study metric on $\cp^1$.

Note that starting with a non-positive quadratic form $\LL'= |x^1|^2 - |x^2|^2$, 
(corresponding to ($\eta_1=1, \eta_2=-1$)), and taking a $U(1)$ quotient leads to
$\LL' = c\ln ( 1 - |z|^2)$,
which gives a Riemannian metric on the hyperbolic disc $|z|<1$.

Notice that the $V$-equations to be solved may become quite complicated if the charges of the fields differ significantly. 

The metric is given by the complex Hessian of the K\"ahler potential:
\beq
ds^2 = \partial_i \bar{\partial}_{\bar k} K dz^i d\bar{z}^{\bar{k}}~~.
\eeq
Here and below $z^i, \bar{z}^i$ will refer to complex coordinates on the quotient, whereas $x^i, \bar{x}^i$ will be reserved for complex coordinates on the original (flat) space.

\subsubsection{The symplectic dual}\lab{sec:sympdual}
For a $(2,2)$ supersymmetric theory where all chiral fields $x^i$ enter the Lagrangian only via the combinations~$|x^i|^2$ (such as~(\ref{free})-(\ref{frees})) it is possible to perform a superspace duality to a theory depending on real linear superfields. One starts by writing a parent action
\beq\lab{WLaction}
S = \int d^6z \left( e^W - WL\right)
\eeq
where $L$ is real linear, $W$ is unconstrained, and $d^6z$ stands for integration over two worldsheet coordinates as well as four superspace coordinates. Integrating out $L$ forces $W$ to be written as the real part of a chiral superfield $W = \phi +\bar{\phi}$, which, after a holomorphic coordinate transformation $x=e^\phi$, becomes $e^W = |x|^2$. On the other hand, integrating out the unconstrained superfield $W$ gives the dual action
\beq
S = \int d^6z \left[-L\ln(L)+L \right]
\eeq
For the model at hand, we rewrite $|x^i|^2 = e^{W^i}$ and introduce the new coordinates $L_i$. The Lagrangian is
\bea
F = \sum_i  e^{W^i + \sum\limits_a V_a Q^a_{\;i}} - \sum_i  W^i L_i -\sum_a V_a c^a
\eea
and the equation for $W^i$ is
\bea
W^i = \ln L_i - \sum_a V_a Q^a_{\;i}
\eea
which allows us to eliminate $W^i$ and get
\bea
F = -\sum_i  L_i(\ln L_i - 1) + \sum_i\sum_a  V_a Q^a_{\;i} L_i
-\sum_a V_a c^a
\eea
The quotient in symplectic coordinates is easier to perform than in K\"ahler coordinates since integrating out $V_a$ just leads to a linear equation
\bea\lab{qlcon}
\sum_i  Q^a_{\;i} L_i = c^a
\eea
which however needs some care in solving since $Q^a_{\;i}$ is a rectangular matrix that is not invertible. The general solution can be written 
\bea\lab{gensol}
L_i = \left( \sum_\al P^i_{\;\al}\mu_\al + \sum_a R^i_{\;a}c^a\right)
\eea
where $P^i_{\;\al}$ is a matrix whose columns span the kernel of $Q^a_{\;i}$ so that $Q^a_{\; i} P^i_{\;\al} = 0$ and $R = Q^t (Q Q^t)^{-1}$ is a matrix such that $QR = id$; here $\mu_\al$ are coordinates on the kernel of $Q$.
Notice that if $d^i = \sum_a R^i_a c^a$ then
\beq
c^a = \sum_i Q^a_{\; i} d^i
\eeq

We can finally write the symplectic potential as
\beq\lab{fsympl}
F = - \sum_i  L_i(\mu) (\ln(L_i(\mu))-1)
\eeq
where we use (\ref{gensol}) to write $L$ in terms of $\mu$ and the FI-terms. The metric in the symplectic coordinates $\mu,\te$ is then
\bea\lab{SympMet}
ds^2=-\left(\frac{1}{4}\frac{\pa^2 F}{\pa \mu_\al\pa \mu_\be}d\mu_\al d\mu_\be+\Big(\frac{\pa^2 F}{\pa \mu_\al\pa \mu_\be}\Big)^{\!\!-1}d\te^\al d\te^\be\right)~. 
\eea
where $\te^\alpha$ are ``angular'' coordinates defined as the imaginary parts of the complex coordinates 
\beq\lab{eq:angular}
z^\al=-\frac12\frac{\pa F}{\pa\mu_\al}+i\te^\al~~.
\eeq

\noindent
{\bf Example:} Using our example from the previous section we start with (\ref{exCP1}) and perform the duality to get a Lagrangian for the dual theory
\beq
-L_1(\ln L_1 -1) - L_2(\ln L_2 -1) -V(c-L_1-L_2)
\eeq
Integrating out the gauge field $V$ imposes the constraint
\beq
L_1+L_2 = c
\eeq
which can be solved by introducing a new superfield $\mu$ so that
\bea
L_1 &=& \mu + \frac{c}{2}\\
L_2 &=& -\mu + \frac{c}{2}
\eea
which gives the symplectic potential
\beq
F(\mu) = -\left(\mu+\frac{c}{2} \right) \left(\ln(\mu+\frac{c}{2} -1\right) -\left(-\mu+\frac{c}{2}\right) \left(\ln(-\mu+\frac{c}{2} -1 \right)
\eeq
Using (\ref{SympMet}) the metric becomes
\beq
ds^2 = \frac{1}{4}\frac{c}{\frac{c^2}{4}-\mu^2} d\mu^2 + \frac{1}{c}\left(\frac{c^2}{4}-\mu^2\right) d\theta^2
\eeq
where we again recognize the round metric on the sphere. Here the range of $\mu$ is the segment $[-\frac{c}2, \frac{c}2]$, which is the moment `polytope' of $S^2$.

We have found that knowledge of the charge matrix $Q$ allows us to immediately construct the GLSM in K\"ahler coordinates whereas knowledge of the toric geometry matrix $P$, which is defined to span the kernel of $Q$, allows us to construct the GLSM in symplectic coordinates.

\subsubsection{The Calabi-Yau condition}
In the sigma model,
the condition that the gauge anomalies cancel \cite{Witten} is
\beq \lab{cycond}
\sum_i Q^a_i=0~.
\eeq
The same condition also implies that the FI-parameters do not receive quantum corrections. This is precisely the requirement that the quotient space has zero first Chern class, i.e., that the quotient space is topologically a Calabi-Yau manifold.

A nice way to see this concretely is by starting from the ambient space $\mathbb{C}^K$ with its canonical holomorphic volume form $\Omega = {\rm d} x^1 \wedge \ldots \wedge {\rm d} x^K$. The existence of a nowhere vanishing holomorphic top form is one of the definitions of a Calabi-Yau manifold. One can only induce a nowhere vanishing holomorphic volume form on the quotient if the form is neutral under all the $U(1)$ actions, which is exactly the condition \eqref{cycond}.

The Calabi-Yau condition can also be formulated for the model in symplectic coordinates using the $P$ matrix. Since the columns of $P$ span the kernel of $Q$, the Calabi-Yau condition is satisfied if there is a linear combination of the columns of $P$ that sums up to a vector where all elements are the same.

\subsection{Ricci-flat sigma models}
After performing the K\"ahler quotient, the GLSM becomes a nonlinear \si-model\footnote{Of course, not every NLSM is obtainable this way.}. We call the holomorphic coordinates on the resulting K\"ahler manifold $z^\al$; these are gauge-invariant functions of the $x^i$ coordinates on the GLSM. In superspace, they are themselves chiral superfields. We begin by briefly reviewing some relevant aspects of such models. We recall that in $(2,2)$ superspace, just as for the linear case (\ref{free}), the Lagrange density is the K\"ahler potential $K(z,\cbar z)$ of the target manifold.

\subsubsection{The $\be$-function}
Classically, any \si-model in two dimensions is conformally invariant on the worldsheet, but quantum corrections in general spoil the invariance. In general the target space metric undergoes renormalization where the one-loop $\be$-function is given by
\beq\lab{Ricci}
\be_{\mu\bar{\nu}} (K)~\propto~ -\frac{\partial^2}{\pa z^\mu \bar{\pa}\bar{z}^{\bar\nu}} \ln\!\left(\!\det\!\left[\frac{\pa^2 K}{\pa\vf^\al\pa\bvf^{\cbar \beta}}\right]\right)~~,
\eeq
where the expression in the r.h.s. is 
the Ricci tensor of the target space.  
One finds that the world sheet theory is not Weyl invariant unless this Ricci tensor vanishes\footnote{However, in general there are corrections to the $\be$-function, as well as to the requirement of Ricci-flatness, starting from four loops~\cite{Zanon1, Zanon2, KazakovZanon}.}. To ensure conformal invariance and finiteness of our theory we therefore look for  
models with Ricci flat target space metrics. 

\subsubsection{The Monge-Amp\`ere equation in Kähler coordinates}
Given the definition of the Ricci tensor we see that our metric is Ricci-flat whenever its determinant can be written as a product of a holomorphic and an antiholomorphic function
\beq\lab{mae}
\det\!\left[\frac{\pa^2 K}{\pa\vf^\al\pa\bvf^{\cbar \beta}}\right]
=f(\vf)\cbar f(\bvf)~,
\eeq
which is a special case of Monge-Amp\`ere equation.

For GLSM's, the toric symmetry guarantees that the K\"ahler potential is independent of the imaginary parts of the coordinates $\vf$; this implies that the product
$f\cbar f$ can only be a function of $\vf^\al+\bvf^{\cbar \al}$,  which implies 
\beq\lab{fexp}
f=e^{A_\al\vf^\al}
\eeq
for any real constants $A_\al$.

\vspace{0.3cm}
\noindent{\bf Example:}
Generically, the quotient metric is not Ricci-flat even if the Calabi-Yau condition is satisfied (which implies that a Ricci-flat metric {\em exists}). Our $\mathbb{CP}^1$ example does not satisfy the Calabi-Yau condition but if we extend the model to the space of the canonical line bundle ${\cal O}(-2)$ over $\mathbb{CP}^1$ by choosing the $Q$ matrix
\beq\lab{QbdlCP1}
Q = 
\begin{pNiceArray}{ccc}
1 & 1 & -2
\end{pNiceArray}
\eeq
we see that the Calabi-Yau condition is satisfied. We may now proceed to write down the sigma model and to find the quotient
\beq
\LL = |x^1|^2 e^V + |x^2|^2 e^V + |x^3|^2 e^{-2V} - cV
\eeq
To simplify the calculation we will go to the singular limit where $c=0$. Then the Kähler potential of the quotient becomes (up to an overall irrelevant constant) 
\beq\lab{KCP1}
K = \left[ |z_1|^2(1+|z_2|^2)^2\right]^{\frac13}
\eeq
which does {\em not} give a Ricci-flat metric since the determinant of the Hessian of this $K$ is
\beq\lab{detCP1}
\frac{2}{27}\left[ |z_1|^2(1+|z_2|^2)^2\right]^{-\frac13} = \frac{2}{27} K^{-\frac13}
\eeq
and obviously does not factorize as in~(\ref{mae}).

\subsubsection{The Monge-Amp\`ere equation in symplectic coordinates}
To write the Ricci-flatness condition in symplectic coordinates we use that the Kähler potential and the symplectic potential are related by a Legendre transformation as well as the relation between the Kähler and symplectic coordinates (\ref{eq:angular}) and insert it in the Monge-Amp\`ere equation (\ref{mae}). This gives a relation involving only the symplectic potential written in symplectic coordinates
\beq\lab{MAsymp}
\det\left[-\frac{\pa^2 F}{\pa \mu_\al\pa \mu_\be}\right]=
e^{\sum_\al \left(A_\al \frac{\pa F}{\pa \mu_\al}\right)}~.
\eeq
If this equation is satisfied, the metric is Ricci-flat.

\subsubsection{Insights using T-duality}
We have seen that the transformation to symplectic coordinates was implemented as a superspace duality transform where the (real part) of the chiral superfield was dualized to a real linear superfield. The duality does not change the background geometry but is interpreted as a change of coordinates. Interestingly, for the same theory, when the model depends only on the real part of the chiral superfield, there is another duality available where the real part of the chiral field is dualized to a real part of a {\em twisted chiral} superfield. The dual target space geometry is now interpreted in a different way, it is the T-dual geometry in the sense of Buscher \cite{BuscherA}, which is in general different than the original geometry. Surprisingly however, both dualities boil down to performing a Legendre transform of the original Kähler potential. That means that we have two alternative interpretations of the symplectic potential~(\ref{fsympl}). As $F(\mu)$ we view it as the symplectic potential of the original geometry but as $F(y+\bar{y})$, where $y$ is a twisted chiral superfield, we interpret it, up to an overall minus sign, as the Kähler potential of the T-dual geometry. This gives us an additional handle on the problem at hand.

Buscher \cite{BuscherA} has shown that the vanishing of the $\beta$-function in a theory is preserved by T-duality to one-loop (see also~\cite{Streets}). We take advantage of the fact that it is straightforward to compute the $\beta$-function in the T-dual geometry. A superspace Lagrangian that depends only on twisted chiral superfields is interpreted as {\em minus} the K\"ahler potential of the geometry. In our case, we have
\beq
\tk(y ,\cbar y )=\sum_i L_i(\ln(L_i)-1) ~,~~\non
\eeq
where now
\beq
L_i = \left( \sum_\al P^i_{\;\al}(y_\al +\cbar y_{\cbar\al})+ \sum_a R^i_{\;a}c^a\right)\equiv [\mathbf{P}(y +\cbar y )+\mathbf{R}c]^i
\eeq
using the matrices $\mathbf{P}$ and $\mathbf{R}$ introduced in subsection \ref{sec:sympdual} and not forgetting the FI parameters $c$. This gives (up to irrelevant terms linear in $y +\cbar y $)
\beq\lab{Tdual}
\tk(y ,\cbar y )=\sum_i[\mathbf{P}(y +\cbar y )+\mathbf{R}c]^i\ln[\mathbf{P}(y +\cbar y )+\mathbf{R}c]^i~~.
\eeq
Taking into account the minus sign in the definition of the K\"ahler potential, the one-loop counterterm is proportional to
\beq\lab{tdualbeta}
\ln\!\Bigg(\!\!\det\!\Bigg[-\frac{\pa^2\tk}{\pa y_\al\pa\cbar y_{\cbar\beta}}\Bigg]\Bigg)=-\ln\!\Bigg(\!\!\det{\sum_i P^i{}_\al\frac1{L_i}P^i{}_{\beta} }\Bigg)~.
\eeq
where the T-dual line element is
\bea
ds^2=\frac{\pa^2\tk}{\pa y_\al\pa\cbar y_{\cbar\beta}}dy_\al d\cbar y_{\cbar\beta}~.
\eea
Thus the Monge-Amp\`ere equation for the T-dual geometry is 
\beq
\det\!\Bigg[ \frac{\pa^2\tk}{\pa y_\al\pa\cbar y_{\cbar\beta}}\Bigg]=\det\left[{\sum_i P^i{}_\al\frac1{L_i}P^i{}_{\beta} }\right] ~\propto~ e^{\sum_\al\widetilde A_\al(y_\al+\cbar y_{\cbar\al})}\,,
\eeq
with $\widetilde A_\al$ constant. 
However, 
\beq
\frac{\pa^2\tk}{\pa y_\al\pa\cbar y_{\cbar\beta}}= - \frac{\pa^2 F}{\pa \mu_\al\pa \mu_\be}~,
\eeq
which implies that the Monge-Amp\`ere equation for the T-dual geometry is precisely the Monge-Amp\`ere equation (\ref{MAsymp})
of the original geometry in symplectic coordinates. 

\subsubsection{Motivation for shadow coordinates/fields}
Here we review the path that led us to quotients of indefinite signature spaces. Recall that we use the terms `shadow coordinates' and `shadow fields' for the negative signature coordinates  and corresponding superfields. 

Let us return to the example of the sigma model defined by the charge matrix (\ref{QbdlCP1}). We saw that the determinant (\ref{detCP1}) is a power of the K\"ahler potential (\ref{KCP1}). This suggests that we might consider some non-trivial power $K^ \al$ of the original Kähler potential $K = \left[ |z_1|^2(1+|z_2|^2)^2\right]^{\frac13}$ and see if we can find a value $\al$ that gives a Ricci-flat metric. Indeed, a quick calculation reveals that $\al=\frac32$ does just that, that is
\beq\lab{CYsingcon}
K_{CY}= \left[ |z_1|^2(1+|z_2|^2)^2\right]^{\frac12}
\eeq
gives a Ricci-flat metric\footnote{In this particular case the metric is not only Ricci-flat but just flat. The corresponding orbifold is $\mathbb{C}^2/\mathbb{Z}_2$, with complex coordinate $u=z^{1/2}$, subject to the identification $u\sim -u$. In more complicated cases even the singular cone will only be Ricci-flat, though.}. The question arises whether we can produce this as a GLSM. The first step was to note that if we write\footnote{If we try $K_{U}=Ke^{U}+|x|^2e^{qU}$, we can integrate out $U$ only if $q$ is negative and the resulting exponent $\al<1$, and thus cannot give the Ricci-flat metric.}
\beq\lab{KVsingen}
K_{U}=Ke^{U}-|x|^2 e^{qU}~~,
\eeq
integrating out $U$ leads to
\beq
K_{q}= \left(\frac{1}{|x|^2}\right)^{\frac{1}{q-1}} (q-1)\left(\frac{K}q\right)^{\frac{q}{q-1}}~~.
\eeq
which after gauge fixing $x=1$ and choosing $q = 3$ gives us the Ricci-flat Kähler potential.
So up to an overall scale, we can realize the Ricci-flat metric as a Kähler quotient of a gauged linear sigma model with charge matrix
\beq
Q = \begin{pNiceArray}{ccc|c}
1 & 1 & -2 & 0\\
1 & 1 & 1 & 3
\end{pNiceArray}
\eeq 
The last column corresponds to the new `shadow' field $x$ which enters with the opposite sign. We separate it from the physical fields with a solid line in the charge matrix. 

By including shadow fields in our construction we need to generalize our formalism. We define a sign factor $\eta_i = \pm 1$  where the plus sign is for physical fields and the minus sign for shadow fields. The linear sigma model we started with is then written as in~(\ref{frees}). The symplectic potential, now with FI parameters turned on, becomes
\bea\lab{gFsympl}
F = -\sum_i  \eta_i L_i(\ln L_i - 1) + \sum_i\sum_a  \eta_i V_a Q^a_{\;i} L_i
-\sum_a V_a c^a
\eea
Just as in section~\ref{sec:sympdual}, varying w.r.t.\,\,the gauge fields $V_a$ gives a system of linear equations on the $L_i$'s. Their solution leads to a modification of the relations (\ref{gensol}) between $L_i$ and~$\mu_\alpha$:
\bea\lab{ggensol}
L_i &=& \eta_i\left( \sum_\al P^i_{\;\al}\mu_\al + \sum_a R^i_{\;a}c^a\right)\equiv \eta_i\left( \sum_\al P^i_{\;\al}\mu_\al + d^i\right)
\eea
The solution is not quite unique; if we shift $\mu_\alpha\rightarrow \mu_\alpha+\de\mu_\alpha$ with some arbitrary constants $\de\mu_\alpha$ we still have a solution but $d^i$ changes:
\bea
d^i = \sum_a R^i_a c^a +\sum_\alpha P^i_\alpha \de\mu_\alpha
\eea
To disentangle this we introduce the matrix $S\equiv (P^tP)^{-1}P^t$
which satisfies $SP=id,~SR=0$ which allows us to separate the contributions
\bea
\lab{cqd}
c^a &=& \sum_i Q^a_{\; i} d^i~~~,~~ \de\mu_\al=\sum_i S_{\al i}d^i~~,
\eea
This gives the final symplectic potential
\bea\lab{finsymp}
F=-\sum_i\left(\sum_\al P^i_{\;\al}\mu_\al + d^i\right)\left[\ln\left(\sum_\al P^i_{\;\al}\mu_\al + d^i\right)-1\right]~.
\eea
Note that the explicit signs $\eta_i$ have dropped out of the expression--the shadow fields are encoded in the matrix $P$. Alternatively, one may keep $L_i$ as in (\ref{ggensol}) which changes the sign of the shadow fields in $P$ (and of the corresponding terms in $d^i$) but then one also needs to keep the $\eta$ in (\ref{gFsympl})

Crucially, the Calabi-Yau condition on the charge matrix needs to be modified in that the shadow fields are counted with opposite sign:
\beq\lab{etaCYcond}
\sum_i \eta_i Q_{\;i}^a = 0\,,
\eeq
which will be derived in section \ref{ssshadow}.
  
Thus we are led to consider quotients of indefinite signature spaces. We shall see below that these shadow coordinates/fields are a generic feature of Ricci-flat GLSMs.
 
\section{Aspects of toric geometry}\lab{fans}
Here we introduce the background knowledge about toric geometry necessary for the discussion that follows.
For more background material on K\"ahler toric geometry see~\cite{Howe,Batyrev,Guillemin,MorrisonTor,Abreu}.
\subsection{Fans}
A toric variety of complex dimension $n$ is a complex algebraic variety that contains the algebraic torus $(\mathbb{C}^\ast)^n$ as a dense open subset. In other words, a toric variety  is obtained by appending the loci of toric degenerations; this additional data -- encoded in a combinatorial data of the fan -- is what distinguishes toric varieties of the same dimension.

A useful set of examples to keep in mind arises in complex dimension one. The three basic toric varieties are $\mathbb{C}^\ast$ (the cylinder), $\mathbb{C}$ (the complex plane), and $\mathbb{P}^1$ (the Riemann sphere). These are obtained from $\mathbb{C}^\ast$ by adding zero, one, or two points, respectively.

An efficient way to encode the combinatorial data defining a toric variety is through a Fan.\footnote{Other equivalent descriptions of toric varieties exist; however, the Fan plays a central role in their relation to gauged linear sigma models (GLSMs).} To define a Fan we let $N = \mathbb{Z}^n$ be an $n$-dimensional integral lattice and $N_{\bbr} := N \otimes \bbr$ be a real vector space. A {\bf convex polyhedral cone} on $N_\bbr$ is defined by\footnote{Later in the text we will only consider convex polyhedral cones and refer to them as cones without any adjectives.} 
\beq\lab{Pmatrix}
    \sigma = \{ a^1 e_1+ \ldots + a^k e_k | a^i > 0 \}~,~~ e_i\in N~,~
     \text{ such that} \;\; \sigma\cap (-\sigma) = \{0\}~~.
\eeq
Then we define a {\bf Fan} as a collection of cones such that any intersection of two cones in the Fan is also a cone. In general, a fan contains cones of dimensions ranging from $1$ up to $n$, together with incidence relations specifying how these cones fit together. Cones of co-dimension one we will refer to as \emph{faces} and the one–dimensional cones will be called \emph{edges}. Note that in complex dimension two, edges and faces coincide in the fan.

Given a fan with $m$ edges in an $n$-dimensional lattice $N$, we define an $m \times n$ matrix $P$ whose rows are the lattice vectors generating the edges of the fan.
\beq
    P = \begin{pmatrix}
        e_1 \\ \vdots \\ e_m
    \end{pmatrix}
\eeq
The data of a fan contains substantially more information than is encoded in the matrix~\(P\) alone. In particular, by specifying only the matrix \(P\) we lose the information about the cone structure and their intersections, retaining only the data of the one–dimensional cones (the edges) of the Fan. Since this reduced set of data is sufficient for the construction of the associated GLSM, we will adopt this simplified description.

In most cases the number of rows of the matrix $P$ greatly exceeds the dimension of $N_{\bbr}$ so the vectors $\{ e_i\}$ are linearly dependent and satisfy linear relations. If there are $k<m$ such relations we encode them in a $k\times m$ matrix $Q$ such that
\beq
    QP = 0
\eeq
Note that both the $P$ and the $Q$ matrix are far from unique. Any row operations on $Q$ give a new $Q$ with an equivalent set of linear relations between the rows of $P$. Similarly any column operations on $P$ gives an equivalent $P$ corresponding to a linear change of basis of~$N_{\bbr}$. Notice that the Calabi-Yau condition (\ref{cycond}) can be easily read off from the $P$ matrix. If there exist a linear combination of the columns of $P$ that gives a column of only $1$'s, then the sum of the charges of each row of $Q$ must be zero.

To illustrate the concept we give an example of a two dimensional toric variety, corresponding to the total space of the $\mathcal{O}(-2)$ bundle over $\cp^1$. Consider a fan:

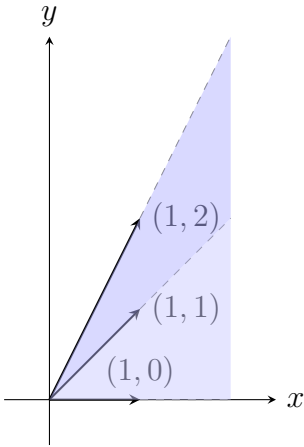
\begin{wrapfigure}{l}{4.5cm}
\vspace{-0.8cm}
\begin{tikzpicture}[scale=1.2, >=stealth]
  % Axes
  \draw[->] (-0.5,0) -- (2.5,0) node[right] {$x$};
  \draw[->] (0,-0.5) -- (0,4) node[above] {$y$};
  % Vectors
  \draw[->, thick] (0,0) -- (1,0) node[right,above] {$(1,0)$};
  \draw[->, thick] (0,0) -- (1,1) node[right] {$(1,1)$};
  \draw[->, thick] (0,0) -- (1,2) node[right] {$(1,2)$};  
    % Extensions from vector tips
  \draw[dashed, gray] (1,0) -- (2,0);
  \draw[dashed, gray] (1,1) -- (2,2);
  \draw[dashed, gray] (1,2) -- (2,4);  
    % Shaded area between (0,1) and (1,1)
  \fill[blue!20, opacity=0.5] (0,0) -- (2,0) -- (2,2) -- cycle;
    \fill[blue!30, opacity=0.5] (0,0) -- (2,2) -- (2,4) -- cycle;
\end{tikzpicture}
\vspace{-1cm}
\end{wrapfigure}
It consists of the following data: two 2-dimensional cones -- shaded regions -- convexly generated respectively by the vectors $(1,0),(1,1)$ and $(1,1), (1,2)$ and one dimensional faces (edges) spanned by the vectors $(1,0),(1,1),(1,2)$.
The intersection of the two cones of dimension $2$ is a face spanned by $(1,1)$ and therefore the conditions on the Fan are satisfied. Although, not fully capturing all the intersection data of the Fan some essential information is captured by matrix $P$ given by
\beq
P  = \begin{pmatrix}
        1 & 0 \\ 1&1 \\ 1&2
    \end{pmatrix}
\eeq
where the rows  correspond to the edges of the fan. Furthermore, since there are more than two vectors there is a relation among the generators encoded in a charge matrix $Q$. In our example it reads
\beq
Q = \begin{pmatrix}
1 &-2 & 1
\end{pmatrix} 
\eeq
which clearly satisfies $QP=0$. Since the first column of $P$ contains only $1$'s, the Calabi-Yau condition is satisfied which can be verified also in $Q$.

Each column of $Q$ gives the charges of one particular chiral superfield in the original GLSM. Then, given a column of $Q$, the corresponding row of $P$ is related to the dual real linear superfield and thus directly tell us how to construct the sigma model in symplectic coordinates.

\subsection{Adding shadows}
Everything in the previous section has a natural extension when the shadow fields are added. Strictly speaking, at the level of the quotient there is no apparent difference between normal and shadow fields. However, in order to make a clear separation we generally list the shadow fields in the $P$ matrix at the bottom and separate them by a line
\begin{equation}
\begin{pmatrix}
P_{\text{normal}} \\ \hline P_{\text{shadow}}
\end{pmatrix}    
\end{equation}
and similarly for the charge matrix where the following form can always be achieved: 
\begin{equation}  
\begin{pNiceArray}{c | c}
Q & 0 \\
\hline
q_1 & q_2
\end{pNiceArray}
\end{equation}
where $Q$ is the original charge matrix of a fan without shadow fields while $q_i$ encode the way the shadows and the original fields are connected.

When adding the shadow fields to $P$, in order to preserve the relation $QP = 0$, we insert them with the opposite sign relative to the physical fields. This sign is canceled when expressing $L_i$ in terms of $\mu_\alpha$ by the $\eta_i$ in (\ref{ggensol}). When constructing either the GLSM from $Q$ or the symplectic sigma model from $P$, we must also remember that  the sign in front of the kinetic terms of the shadow fields must also be opposite to the physical ones.

\section{Generalizing the Calabi Ansatz}\lab{CalAn}
In this section we begin by reviewing the Calabi ansatz construction \cite{Calabi1979} as a method for constructing a Ricci flat metric on a particular kind of geometry -- a complex cone $\mcl{C}$ over a K\"ahler-Einstein manifold $\mcl{B}$ with complex dimension $D-1$. These manifolds are examples of non-compact Calabi-Yau manifolds, asymptotic to (real/metric) cones over Sasaki-Einstein manifolds (see~\cite{SparksSasaki} for more on this). An extension of the Calabi-Yau theorem for such manifolds, asserting the existence of Ricci-flat metric in every K\"ahler class, was proven in~\cite{vanCoeveringGeneral, Goto}. 

Calabi's ansatz leads to explicit expressions in the case when the base $\mathcal{B}$ is a homogeneous space, cf.~\cite{GibbonsPope}, \cite{CdO}, \cite{Nitta1}, \cite{Nitta2}, \cite{Nitta3}, \cite{Nitta4},
\cite{CorreaFlags}. A generalization of Calabi's ansatz that will be described in section \ref{gencal} allows constructing Ricci-flat metrics in \textit{every K\"ahler class} for a large family of examples\footnote{We should note that there are methods for constructing Ricci-flat metrics that go beyond the (generalized) Calabi's ansatz by exploiting conformal Killing-Yano tensors~\cite{Gauduchon},  \cite{Semmelmann}, \cite{ChenLuPope}, 
\cite{Martelli:2007pv}, \cite{Bykov2017}, \cite{Ovchinnikov}. These will not be relevant for us in the present paper, however.} (in practice this means that in those cases the metric features more than a single parameter), see also the related work~\cite{Douglas:1997zj}, \cite{Pedersen}, \cite{vanCoevering}, \cite{AZadeBykov}. Utilizing the symplectic point of view, we will provide a GLSM interpretation of such metrics in terms of an extended GLSM model with a  new sector of  shadow~fields.

\subsection{Standard Calabi Ansatz}
A K\"{a}hler-Einstein manifold $\mathcal{B}_{D-1}$ is a K\"{a}hler manifold of complex dimension $D-1$ with a further condition that the Ricci tensor is proportional to the metric 
\beq
{\rm Ric} ~ \propto~ \lambda g\,,\quad\quad \lambda =\mathrm{const.}
\eeq
Suppose now that $\Phi(z,\bvf)$ is a K\"ahler potential of this metric; then the metric obeys the following Monge-Amp\`ere equation:
\beq
\det(\pa\cbar\pa {\Phi})~\propto~ e^{-\la{\Phi}}~,
\eeq
where the coefficient of proportionality may be the norm of any (locally) holomorphic section of a line bundle. 

Now consider a complex cone over $\mcl{B}$ with a K\"ahler potential constructed from the old one as\footnote{The same form occurs in superspace supergravity in 4 spacetime dimensions with the conformal compensator. It might be interesting to see if Ricci-flatness of the space with the conformal compensator has a physical significance.}
\beq \lab{noglsm}
K_\mcl{C}=z\cbar z e^{\al{\Phi}}
\eeq
with the $z$ coordinate being the complex coordinate along the cone direction. Upon computing the K\"ahler form  we find
\beq
\pa\cbar\pa K_\mcl{C} = K_\mcl{C}\left(\al \pa\cbar\pa{\Phi} +\al^2 \pa{\Phi}\cbar\pa{\Phi}+\al\pa{\Phi} \frac{d\cbar z}{\cbar z}+
\al\frac{dz}{z}\cbar\pa{\Phi}+\frac{dzd\cbar z}{z\cbar z}\right)~.
\eeq
Up to powers of $|z|^2$, which we may ignore when we calculate the Ricci-form, the determinant of this is
\beq
\det(\pa\cbar\pa K_\mcl{C})~\propto ~K_\mcl{C}^D \det{\pa\cbar\pa{\Phi}}~\propto~ e^{\al D{\Phi}}e^{-\lambda{\Phi}}~~,
\eeq
which means that $K_\mcl{C}$ gives a Ricci-flat metric when $\la=\al D$, or
\beq\lab{al}
\al=\frac\la{D}~.
\eeq

\subsubsection{Calabi Ansatz in symplectic coordinates}
Clearly \eqref{noglsm} is not in the simple form of a GLSM. However, performing a Legendre transform in the $z$ coordinate -- the cone direction -- we obtain
\beq
F_\mcl{C} = e^{W+\al \Phi}-\mu W \then W=\ln\mu-\al\Phi ~~,
\eeq
and hence
\beq
F_\mcl{C}=-\mu(\ln\mu-1-\al\Phi)
\eeq
If $\mcl{B}_{D-1}$ itself is toric, we find that the full  symplectic potential is given by a further Legendre transform
\beq
F=-\mu\left(\ln\mu-1-\al{\Phi}(W^i)\right)-\mu_iW^i \then
\al\mu\frac{\pa{\Phi}}{\pa W^i}=\mu_i~~,
\eeq
which gives
\beq\lab{CA_sym}
F= -\mu(\ln\mu-1) +\al\mu F_\Phi\left(\frac{\mu_i}{\al\mu}\right)~~,
\eeq
where $F_\Phi$ is the usual Legendre transform of $\Phi$, i.e., the
symplectic potential for $\mcl{B}$. In general, the symplectic potential (\ref{CA_sym}) is rather complicated, but as we shall see below,  it simplifies in special cases.
\subsubsection{Application to generalized singular conifolds.}
We can directly apply these general results. We start with a product of $N$ projective spaces
\beq
\mcl{B}_{\cp}=\prod\limits_{A=1}^N\,\cp^{\nsa-1}
\eeq
and construct a K\"ahler-Einstein metric on it. The first step is to consider the Fubini-Study metric on each factor with arbitrary radii~$\be^A$: 
\beq\lab{FSP}
\Phi=\sum_{A=1}^N \be^A\ln(|x_A|^2)~~,~~ 
|x_A|^2\equiv\sum_{i=1}^{\nsa} x^{i}_A \cbar x^{i}_A~~,
\eeq
where $x^{i}_A,~i=1\;...\;\nsa$ are $\nsa$ homogeneous coordinates on $\cp^{\nsa-1}$. The dimension of~$\mcl{B}_{\cp}$ is 
\beq
D-1\equiv\Big(\sum_A \nsa\Big)-N~.
\eeq
Evaluating the determinant, we find
\beq
\det(\pa\cbar\pa\Phi)~\propto~ \prod_{A}^N \left(|x_A|^2\right)^{-\nsa}~.
\eeq
Comparing this to $e^{-\la\Phi}$, we find 
\beq\lab{betaval}
\be^A=\frac{\nsa}\la~;
\eeq
Now using the result for the Calabi ansatz (\ref{al}), we find 
\beq
\al\be^A=\frac{\nsa}D~.
\eeq
\subsubsection{Metric on   
$\mathcal{B}_{\cp}$ in symplectic coordinates}
From the K\"ahler potential~(\ref{FSP}) for the Fubini-Study metric on the product  with coefficients given by (\ref{betaval}), we construct a dual potential depending on unconstrained real fields $W_A^i$ and real linear fields $L^A_i$:
\beq
F_\Phi=\sum_A\left(\frac{\nsa}\la\ln\left(\sum_i^{\nsa}e^{W_A^i}\right)-\sum_i^{\nsa}\hat L^A_iW^i_A\right)\,,
\eeq
similarly to what we did for flat space in~(\ref{WLaction}). Extremization w.r.t. $W_A^i$  gives
\beq
\frac{\nsa}\la\,\frac{e^{W_A^i}}{\sum\limits_i^{\nsa}e^{W_A^i}}=\hat L^A_i\then \frac{\nsa}\la=\sum_i^{\nsa}\hat L^A_i~.
\eeq
Eliminating $W^i_A$ amounts to a Legendre transform. Modulo irrelevant terms linear in $\hat L^A_i$, we find
\beq\lab{FproductCPN}
F_\Phi = -\sum_A^N\sum_i^{\nsa}\hat L^A_i\ln(\hat L^A_i)~~~~\hbox{with}~~~~
\frac{\nsa}\la=\sum_i^{\nsa}\hat L^A_i~.
\eeq
We can solve the constraint, e.g., by writing
\beq\lab{lmusol}
\hat L_i^A=\mu_i^A+\frac1\la~~\hbox{for}~~i<\nsa~~,~~ \hat L_{\nsa}^A=\frac1\la-\sum_i^{\nsa-1}\mu_i^A~~.
\eeq
\subsubsection{Metric on the cone over 
$\mathcal{B}_{\cp}$ in symplectic coordinates}
We now use (\ref{CA_sym}) to find the symplectic potential for the Ricci-flat complex cone over~$\mathcal{B}_{\cp}$. The geometric interpretation of this space is that it is the total space of the canonical bundle
\beq
\mcl{K}_{\mcl{B}_{\cp}}=\otimes_{A=1}^N\,\mcl{O}(-\nsa)
\eeq
with the zero section contracted to zero. In the next section we will discuss the smooth Ricci flat metric on the total space of $\mcl{K}_{\mcl{B}_\cp}$ itself.

Substituting~(\ref{FproductCPN}) in~(\ref{CA_sym}), we have
\beq
F= -\mu(\ln\mu-1)+\al\mu\sum_A^N\left[-\sum_i^{\nsa}\frac{L^A_i}{\al\mu}(\ln(L^A_i)-\ln(\al\mu))+\ln\left(\frac{\nsa}\la\right)\frac{L^A_i}{\al\mu}\right]
\eeq
with
\beq
\frac{\nsa}\la=\sum_i^{\nsa}\frac{L^A_i}{\al\mu}~.
\eeq
Simplifying and dropping constant and linear terms, we find
\beq
F=-\mu\ln\mu +\sum_A^N\left[-\sum_i^{\nsa}L^A_i\ln(L^A_i)+\frac{\nsa}D\mu\ln\mu\right]~~~~\hbox{with}~~~~
\sum_i^{\nsa}L^A_i=\frac{\nsa}D\mu~.
\eeq
The $D$ in the denominator does not have a toric interpretation (see section \ref{fans}), so we must rescale $\mu$ to remove it. This gives the final symplectic potential of the singular generalized conifold (again dropping linear and constant terms):
\beq\lab{Fsingcon}
F=-D\mu\ln\mu +\sum_A^N\left[-\sum_i^{\nsa}L^A_i\ln(L^A_i)+\nsa\,\mu\ln\mu\right]~~~~\hbox{with}~~~~
\sum_i^{\nsa}L^A_i=\nsa\,\mu~.
\eeq
As in (\ref{lmusol}), we can solve the constraint on the $L^A_i$ in terms of unconstrained symplectic coordinates $\mu^A_i$ by
\beq\lab{lmusolcone}
L_i^A=\mu_i^A+\mu~~\hbox{for}~~i<\nsa~~,~~ L_{\nsa}^A=\mu -\sum_i^{\nsa-1}\mu_i^A~~.
\eeq
\subsection{The generalized Calabi ansatz -- resolved case} \lab{gencal}
In the previous section we gave a full symplectic potential yielding a singular Ricci flat metric on the cone over a product of projective spaces. The ansatz extends beyond the singular case with small modification.

The resolved symplectic potential is  obtained from  \eqref{Fsingcon} by introducing shift parameters into the constraints and by adding an extra function $F_R(\mu)$ 
\begin{gather} \lab{F_ansatz}
F_G=-F_R(\mu) -\sum_A^N\left[\sum_i^{\nsa}\Big(L^A_i\ln(L^A_i)- s^A\ln s^A\Big)\right]
\\
~~\hbox{with}~~
\sum_i^{\nsa}L^A_i=\nsa(\mu+a_A)~\equiv \nsa\,s^A~~.
\end{gather}
There is a limit which reduces the resolved ansatz back to \eqref{Fsingcon} -- modulo a linear term -- when $a_A=0$ and the extra function is set to $F_R=D\mu\ln\mu$.

In the resolved case $F_R$ is fixed by imposing the Monge-Amp\`ere equation \eqref{MAsymp}. Since $F_R$ enters the equation only through its derivatives, it is handy to express it in terms of another auxiliary function~$R(\nu)$ as
\beq
F_R(\mu) =\int^\mu \ln(R(\nu))\,d\nu~~~.
\eeq
After some straightforward but tedious calculations, detailed in Appendix~\ref{MAapp}, the Monge-Amp\`ere equation reduces to
\beq\lab{diffR}
R'=\prod_A^N\frac{(s^A)^{\nsa-1}}\nsa~~.
\eeq
The singular limit $a_A=0$ indeed reproduces $F_R=D\mu\ln\mu$.

Since $R'$ is a polynomial in $\mu$, so is $R$, and hence $F_R$ is
a sum of terms 
\beq
F_R=-\sum_1^D L_I\ln(L_I)~~\hbox{where}~~R(\mu)=\prod_1^D L_I~~.
\eeq
This allows us to read off the $P$ matrix and thus find the gauged linear sigma model as discussed above and as will be elaborated  in section~\ref{toricdata} below.

It is often convenient to put $R$ into ``depressed form'',
 that is, shifting $\mu$ to cancel the second highest power of $\mu$; clearly, then $R'$ is automatically also in depressed form. Another ``standard" form is to choose one of the roots of $R'$ to be at the origin; then $R$ is missing the linear term.

\subsubsection{Examples} 
\paragraph{Cone over $\cp^{n-1}$:}
The simplest example starts with a $\mcl{B}$ just a single $\cp^{n-1}$~\cite{Pedersen}; here $n=D$ and we have
\beq
s=\mu~\then~R'=\frac1n\mu^{n-1} ~\then ~ R=\frac1{n^2}(\mu^{n}-b^n)
\eeq
where we have shifted $\mu$ to put $R,R'$ into depressed form,  and the term with $b$ is an arbitrary integration constant. This can be factorized as 
\beq
R= \frac1{n^2}\prod_{i=0}^{n-1}(\mu-b\om^i)~~,
\eeq
where $\om^n=1$. This gives (modulo constant and linear terms that we drop)
\beq
F_R= \sum_{i=0}^{n-1}(\mu -b\om^i)\ln(\mu -b\om^i)~,
\eeq
and the complete symplectic potential
\bea\lab{fp2}
F&=& \sum_{i=0}^{n-1}\;[\mu\ln\mu-(\mu -b\om^i)\ln(\mu -b\om^i)]-
\sum_{i=1}^{n-1}(\mu_i+\mu)\ln(\mu_i+\mu)\non[1mm]
&&-~\left(\mu-\sum_{i=1}^{n-1}\mu_i\right)\ln\left(\mu-\sum_{i=1}^{n-1}\mu_i\right)
~.
\eea
Note the curious feature that though $F$ is perfectly real, for $n>2$, it involves complex numbers.  
\paragraph{Cone over $\cp^1\times\cp^1$:}
Another example that we examine in detail is $\mcl{B}=\cp^1\times\cp^1$~\cite{CdO, PandoZayasTseytlin}. In this case, $N=n_1=n_2=2$ and $D=3$. We choose $R'=\frac14\mu(\mu+a)$; then the roots of $R$ obey
\beq\lab{abrel}
-\frac23\sum_{I=1}^3b_I=a ~~\hbox{and}~~\sum_{I=1}^3\frac1{b_I}=0
\eeq
and we find
\bea\lab{fp1p1}
F&=&-\left[\sum_{I=1}^3 (\mu-b_I)\ln(\mu-b_I)\right] +2 \mu\ln\mu+2(\mu+a)\ln(\mu+a)\non[2mm]
&&
-(\mu_1+\mu+a)\ln(\mu_1+\mu+a)
-(\mu+a-\mu_1)\ln(\mu+a-\mu_1)\non[2mm]
&&
-(\mu+\mu_2)\ln(\mu+\mu_2)-(\mu-\mu_2)\ln(\mu-\mu_2)~~.
\eea
As a result of~(\ref{abrel}) there are two genuine parameters (the K\"ahler moduli), which may be thought of as the radii of the two $\cp^1$'s specifying the K\"ahler class of the metric within $H^2(\cp^1\times\cp^1, \bbr)=\bbr^2$.

\subsection{Reading off the toric data} \lab{toricdata}
From (\ref{Fsingcon}), the symplectic potential in the singular limit, the $P$ matrix of the model is most easily read off by comparing it to (\ref{gFsympl}) using the identification (\ref{ggensol}). By construction, for each value of $A$ the $L^A_i$ is built from the $P$ matrix of the $\mathbb{CP}^{n_A-1}$ in question. If we call the relevant data $\hat{L}^A$ and $\hat{P}^A$, then
\beq
\hat{L}^A_i(\mu^A) = \sum_\alpha(\hat{P}^A)^{i}_{\;\;\alpha} \mu^A_\alpha
\eeq
However, for $\mathbb{CP}^{n_A-1}$ we know that $\sum_i^{n_A} \hat{L}^A_i = 0$ which is in disagreement with the constraint in (\ref{Fsingcon}). 
To solve this we introduce the additional coordinate $\mu$ such that for each value of $A$ we have
\bea
L^A_i &=& \hat{L}^A_i(\mu_A) + \mu \\
L &=& \mu
\eea
which satisfies $\sum L^A_i = n_A L$. This corresponds to adding an additional column of ones into the $P$ matrix as well as a row of zeros with a $1$ in the final column representing $L$.

\begin{wrapfigure}{r}{6cm}
\vspace{-0.8cm}
     \beq\lab{Pwrapmat1}
\begin{pNiceArray}{cIcIcIc}
\hat{P}^1 & 0 & 0 & 1\\
\hdashline
0 & \ddots & 0 & \vdots\\
\hdashline
 0 & 0 & \hat{P}^N & 1\\
 \hdashline
 0 & 0 & 0 & 1
\end{pNiceArray}
\eeq
\vspace{-0.8cm}
\end{wrapfigure}

The $P$ matrix constructed so far can be written as in~(\ref{Pwrapmat1}). 
To this we need to append rows corresponding to the terms $-D\mu\ln \mu$ and $\sum_A^{N} n_A \mu \ln \mu$ (using that $L = \mu$) which combines to $(N-1)\mu\ln\mu$ since in the singular limit these two terms can cancel. \begin{wrapfigure}{r}{6cm}
\vspace{-1cm}
\beq\lab{Pwrapmat2}
\begin{pNiceArray}{cIcIcIc}
            \hat{P}^1 & 0 & 0 & 1\\
            \hdashline
            0 & \ddots & 0 & \vdots\\
            \hdashline
            0 & 0 & \hat{P}^N & 1 \\
            \hdashline
            0 & 0 & 0 & 1\\
            \hline
            0 & 0 & 0 & -1\\
            \vdots & \vdots & \vdots & \vdots \\
            0 & 0 & 0 & -1\\         
        \CodeAfter
            \SubMatrix{.}{5-1}{7-4}{{\} N}}[xshift=3.5mm]
        \end{pNiceArray}
\eeq    
\end{wrapfigure}
To find the appropriate number of terms to keep we need to look at the resolved case. From (\ref{F_ansatz}) we see that for each $\mathbb{CP}^{n_A-1}$ factor a deformation parameter is added to the positive sign term  which means that in the singular limit we must have
\beq
-\mu\ln\mu + N \mu\ln\mu
\eeq

The first term is already encoded in the row with only the last entry of $1$ being nonzero. The second term is more interesting. It comes with the ``wrong'' sign and thus corresponds to our shadow fields. In the $P$ matrix they correspond to rows with the last column equal to $-1$. The final form of the $P$ matrix is the $(2N+D)\times D$ matrix shown in~(\ref{Pwrapmat2}), where the solid line divides the shadow sector from the rest.
This gives us the charge matrix $Q$ of the GLSM as a $2N\times (2N+D)$ matrix
\beq
Q =
\begin{pNiceArray}{ccccc|cccc}[first-row]
        \rule[-8pt]{0pt}{8pt}
        \Block{1-1}{\vspace{-0.3cm}\overbrace{}^{\displaystyle n_1}} 
        & \Block{1-1}{\vspace{-0.3cm}\overbrace{}^{\displaystyle n_2}}
        & & 
        \Block{1-1}{\vspace{-0.3cm}\overbrace{}^{\displaystyle n_N}}
        & &  &\Block{1-2}{\vspace{-0.3cm}\overbrace{\hspace{2.2cm}}^{\displaystyle N}}\\
            1\cdots 1 & 0\cdots 0 & \cdots & 0\cdots 0 & -n_1 & 0 & 0 & \cdots & 0\\
%            \hline
            0\cdots 0 & 1\cdots 1 & \cdots & 0\cdots 0 & -n_2 & 0 & 0 & \cdots & 0\\
%            \hline
            0\cdots 0 & 0\cdots 0 & \ddots & 0\cdots 0 & \vdots & \vdots &  \vdots & \ddots & \vdots \\
%            \hline
            0\cdots 0 & 0\cdots 0 & \cdots & 1\cdots 1 & -n_N & 0 & 0 &\cdots & 0\\
            \hline
            0\cdots 0 & 0\cdots 0 & \cdots & 0\cdots 0 & 1 & 1 & 0 &\cdots & 0\\
            0\cdots 0 & 0\cdots 0 & \cdots & 0\cdots 0 & 1 & 0 & 1 &\cdots & 0\\
            \vdots & \vdots & \ddots & \vdots & \vdots & \vdots & \vdots & \ddots & \vdots \\
            0 \cdots 0 & 0\cdots 0 & \cdots & 0\cdots 0 & 1 & 0 & 0 &\cdots & 1\\
        \CodeAfter
             \SubMatrix{.}{1-1}{4-9}{{\} N}}[xshift=3.5mm]
             \SubMatrix{.}{5-1}{8-9}{{\} N}}[xshift=3.5mm]
        \end{pNiceArray}
\eeq

Going away from the singular limit we will add deformation parameters to the shadow fields
\beq
s^A = \mu \rightarrow s^A = \mu + a^A
\eeq
so these terms cannot cancel with the terms from $F_R$ anymore. 

\begin{wrapfigure}{r}{7.5cm}
\vspace{-0.3cm}
\beq\lab{Pfull} \hspace{-2cm}
P =
\begin{pNiceArray}{cIcIcIc}
            \hat{P}^1 & 0 & 0 & 1\\
            \hdashline
            0 & \ddots & 0 & \vdots\\
            \hdashline
            0 & 0 & \hat{P}^N & 1 \\
            \hdashline
            0 & 0 & 0 & 1\\
            \vdots & \vdots & \vdots & \vdots \\
            0 & 0 & 0 & 1\\
            \hline
            0 & 0 & 0 & -1\\
            \vdots & \vdots & \vdots & \vdots \\
            0 & 0 & 0 & -1\\         
        \CodeAfter
            \SubMatrix{.}{4-1}{6-4}{{\} D}}[xshift=3.5mm]
            \SubMatrix{.}{7-1}{9-4}{{\} N+D-1}}[xshift=3.5mm]
        \end{pNiceArray}
\eeq
\vspace{-1.5cm}
\end{wrapfigure} 
For each factor in the base we add $n_A$ shadow fields giving the contribution to the symplectic potential $n_A s^A \ln s^A$ which gives $N+D-1$ shadow fields in total. From the Ricci-flatness  equation~(\ref{diffR}) we find that 
%the metric is Ricci flat if
%\beq\lab{eq:RFmy}
%R' = \prod\limits_A^N \frac{(s^A)^{n_A-1}}{n_A}
%\eeq
%which tells us that 
$R$ is a polynomial in $\mu$ of degree $D$ so there are $D$ physical fields opposed to the shadow fields. The $P$ and $Q$ matrices look the same as in the singular limit although the number of rows of $P$ and columns of $Q$ have increased.

\vspace{0.5cm}
However, one should remember that for each $\mathbb{CP}^{n_A-1}$ factor in the base we have $n_A$ shadow fields that all behave in the same way. For instance, when we deform they are all shifted by the same deformation parameter.

We may now write the $Q$ matrix in a canonical form that we choose to be
\beq\lab{mainQ}
\hspace{-2cm}
Q =
\begin{pNiceArray}{cccc I ccccc | cccc}[first-row]
        \rule[-8pt]{0pt}{8pt}
        \Block{1-1}{\vspace{-0.3cm}\overbrace{}^{\displaystyle n_1}} & \Block{1-1}{\vspace{-0.3cm}\overbrace{}^{\displaystyle n_2}} & &
        \Block{1-1}{\vspace{-0.3cm}\overbrace{}^{\displaystyle n_N}} & \Block{1-5}{\vspace{-0.3cm}\overbrace{\hspace{3.3cm}}^{\displaystyle D}} & &  & & & &\Block{1-2}{\vspace{-0.3cm}\overbrace{\hspace{2.2cm}}^{\displaystyle N+D-1}}\\
            1\cdots 1 & 0\cdots 0 & \cdots & 0\cdots 0 & 
            -n_1 & 0 & 0 & \cdots & 0 & 0 & 0 & \cdots & 0\\
%            \hline
            0\cdots 0 & 1\cdots 1 & \cdots & 0\cdots 0 & 
            -n_2 & 0 & 0 & \dots & 0 & 0 & 0 & \cdots & 0\\
%            \hline
            0\cdots 0 & 0\cdots 0 & \ddots & 0\cdots 0 & 
            \vdots & \vdots & \vdots & \ddots & \vdots & \vdots & \vdots & \ddots & \vdots \\
%            \hline
            0\cdots 0 & 0\cdots 0 & \cdots & 1\cdots 1 & 
            -n_N & 0 & 0 & \cdots & 0 & 0 & 0 &\cdots & 0\\
            \hdashedline
            0\cdots 0 & 0\cdots 0 & \cdots & 0\cdots 0 & -1 & 1 & 0 &\cdots & 0 &
            0 & 0 & \cdots & 0\\
            0\cdots 0 & 0\cdots 0 & \cdots & 0\cdots 0 & -1 & 0 & 1 &\cdots & 0 &
            0 & 0 & \cdots & 0\\
            \vdots & \vdots & \ddots & \vdots & \vdots & \vdots & \vdots & \ddots & \vdots &
            \vdots & \vdots & \ddots & \vdots\\
            0 \cdots 0 & 0\cdots 0 & \cdots & 0\cdots 0 & -1 & 0 & 0 &\cdots & 1 &
            0 & 0 & \cdots & 0\\
            \hline
            0\cdots 0 & 0\cdots 0 & \cdots & 0\cdots 0 & 1 & 0 & 0 &\cdots & 0 &
            1 & 0 & \cdots & 0\\
            0\cdots 0 & 0\cdots 0 & \cdots & 0\cdots 0 & 1 & 0 & 0 &\cdots & 0 &
            0 & 1 & \cdots & 0\\
            \vdots & \vdots & \ddots & \vdots & \vdots & \vdots & \vdots & \ddots & \vdots &
            \vdots & \vdots & \ddots & \vdots\\
            0 \cdots 0 & 0\cdots 0 & \cdots & 0\cdots 0 & 1 & 0 & 0 &\cdots & 0 &
            0 & 0 & \cdots & 1\\
        \CodeAfter
             \SubMatrix{.}{1-1}{4-13}{{\} N}}[xshift=3.5mm]
             \SubMatrix{.}{5-1}{8-13}{{\} D-1}}[xshift=3.5mm]
             \SubMatrix{.}{9-1}{12-13}{{\} N+D-1}}[xshift=3.5mm]
        \end{pNiceArray}
\eeq

The deformation parameters can be collected in a vector $d$ of size $3D+2N-2$. 
If we let $a^A$ be the parameters associated with the shadow fields (one parameter for all shadow fields associated to one particular $\cp^{n_A-1}$) and $b^m$ ($1\leq m \leq D$) the zeros of the polynomial $R(\mu)$, 
\begin{wrapfigure}{r}{4.5cm}
    \beq\lab{vecwrap} \hspace{-0.8cm}
d=\begin{pNiceArray}{c}
            a_1\\
            \vdots\\
            a_1\\
            \hdashline
            %a_2\\
            %\vdots\\
            %a_2\\
            %\hdashline
            \vdots\vdots\\
            \hdashline
            a_N\\
            \vdots\\
            a_N\\
            \hdashline
            b_1\\
            \vdots\\
            b_D\\
            \hline
            -a_1\\
            \vdots\\
            -a_1\\
            \vdots\vdots\\
            -a_N\\
            \vdots\\
            -a_N\\
        \CodeAfter
            \SubMatrix{.}{1-1}{3-1}{{\} n_1}}[xshift=3.5mm]
            %\SubMatrix{.}{4-1}{6-1}{{\} n_2}}[xshift=3.5mm]
            \SubMatrix{.}{5-1}{7-1}{{\} n_N}}[xshift=3.5mm]
            \SubMatrix{.}{8-1}{10-1}{{\} D}}[xshift=3.5mm]
            \SubMatrix{.}{11-1}{13-1}{{\} n_1}}[xshift=3.5mm]
            \SubMatrix{.}{15-1}{17-1}{{\} n_N}}[xshift=3.5mm]
        \end{pNiceArray}
\eeq
\vspace{-3cm}
\end{wrapfigure}
they are related by (\ref{diffR}). 
To this we need to add the appropriate combinations of $a^A$ for the constraints to remain true:
\beq
\sum_i^{n_A} L_i^A = n_A\mu = n_A s^A \rightarrow \sum_i^{n_A} (L_i^A + a^A) = n_A (\mu + a^A)
\eeq

These relations were essential in proving that the Monge-Amp\`ere equation holds. This gives us the vector shown in~(\ref{vecwrap}).

One can check that for our class of models, $\de\mu^\alpha = S_{\alpha i}d^i = 0$ if $\sum b_I = 0$.
Finally, the FI terms in the GLSM (vector $c$) are then given by acting with $Q$ on $d$.

\subsubsection{Examples}
We now apply our prescription to a few nontrivial examples. 
\paragraph{Cone over $\cp^2$:} We start with (\ref{fp2}), and find the following expressions for the $P$ and $Q$ matrices:

\begin{align}
P =
\begin{pNiceArray}[t]{ccc}
1&0&1\\
0&1&1\\
-1&-1&1\\
0&0&1\\
0&0&1\\
0&0&1\\
\hline
0&0&-1\\
0&0&-1\\
0&0&-1\\
\end{pNiceArray}\,,\quad 
\textrm{so that}\quad 
Q =
\begin{pNiceArray}[t]{ccc I ccc | ccc}
1&1&1&-1&-1&-1&0&0&0\\
\hdashline
0&0&0&-1&1&0&0&0&0\\
0&0&0&-1&0&1&0&0&0\\
\hline
0&0&0&1&0&0&1&0&0\\
0&0&0&1&0&0&0&1&0\\
0&0&0&1&0&0&0&0&1 
\end{pNiceArray},
\end{align}
\begin{wrapfigure}{r}{9cm}
\vspace{-1.5cm}
    \beq\lab{dcwrap1}
d=
b\begin{pNiceArray}{c}
0\\
0\\
0\\
1\\
\om\\
\om^2\\
\hline
0\\
0\\
0\\
\end{pNiceArray}~~\quad \Longrightarrow\quad c=
b\begin{pNiceArray}{c}
0\\
\om -1\\
\om^2 -1\\
\hline
1\\
1\\
1\\
\end{pNiceArray}~~.
\eeq
\vspace{-2cm}
\end{wrapfigure}

\vspace{0cm}\noindent
where the latter matrix is related to (\ref{mainQ}) through simple row operations. 
We can also read off the $d^i$ by comparing to (\ref{fsympl}) and then find the FI-terms using~(\ref{cqd}), resulting in~(\ref{dcwrap1}) (here~$\om^3=1$).

\vspace{1cm}
Thus our final GLSM is:
\begin{align}
&K= \sum\limits_1^3 \Big(|x^i|^2e^{V_1}\Big)+|x^4|^2e^{-V_1+\sum\limits_2^6V_A}+
|x^5|^2e^{-V_1-V_2}+|x^6|^2e^{-V_1-V_3}-\\%\non[3mm] 
\nonumber
&\quad\quad\quad\quad\quad\quad\quad\quad\quad\quad\quad\quad-\sum\limits_4^6|x^{3+A}|^2e^{V_A}-b\Big[(1-\om)V_2+(1-\om^2)V_3+\sum\limits_4^6V_A\Big]~.
\end{align}
Note that though the FI terms are complex\footnote{In \cite{Douglas:1997zj} it was pointed out that such terms are needed to get Ricci-flat metrics.}, after we perform the quotient we get a real K\"ahler potential; this is not really a surprise, since $K$ is preserved if we complex conjugate and simultaneously interchange 
\beq
V_2 \lra V_3 ~~, ~~ x^5 \lra x^6~~.
\eeq

\paragraph{Cone over $\cp^1\times\cp^1$:} We start with the symplectic potential~(\ref{fp1p1}) and read off the $P$ and $Q$ matrices (related to the form \eqref{mainQ} by row operations): 
\beq
P =
\begin{pNiceArray}[t]{ccc}
1&0&1\\
0&1&1\\
-1&0&1\\
0&-1&1\\
0&0&1\\
0&0&1\\
0&0&1\\
\hline
0&0&-1\\
0&0&-1\\
0&0&-1\\
0&0&-1\\
\end{pNiceArray}\,,\quad \textrm{so that}\quad 
Q =
\begin{pNiceArray}[t]{cccc I ccc | cccc}
1&-1&1&-1&0&0&0&0&0&0&0\\
1&0&1&0&-2&0&0&0&0&0&0\\
\hdashline
0&0&0&0&-1&1&0&0&0&0&0\\
0&0&0&0&-1&0&1&0&0&0&0\\
\hline
0&0&0&0&1&0&0&1&0&0&0\\
0&0&0&0&1&0&0&0&1&0&0\\
0&0&0&0&1&0&0&0&0&1&0\\
0&0&0&0&1&0&0&0&0&0&1\\
\end{pNiceArray}
\eeq

\begin{wrapfigure}{r}{9cm}
\vspace{-2.5cm}
    \beq\lab{dcwrap2}
d=
\begin{pNiceArray}{c}
a\\
0\\
a\\
0\\
b_1\\
b_2\\
b_3\\
\hline
-a\\
-a\\
0\\
0\\
\end{pNiceArray}~~ \quad \Longrightarrow\quad c=
\begin{pNiceArray}{c}
2a\\
2(a-b_1)\\
b_1-b_2\\
b_1-b_3\\
\hline
b_1+a\\
b_1+a\\
b_1\\
b_1\\
\end{pNiceArray}~~
\eeq
\vspace{-4cm}
\end{wrapfigure}
Again, we read off the $d^i$ by comparing to (\ref{fsympl}) and find the FI-terms using~(\ref{cqd}). The result is shown in~(\ref{dcwrap2}),  
where the parameters are subject to the constraints (\ref{abrel})

\beq\nonumber 
\hspace{-9cm}-\frac23\sum_{I=1}^3b_I=a ~~\hbox{and}~~\sum_{I=1}^3\frac1{b_I}=0~~.
\eeq

\vspace{1cm}
\section{Shadow fields from generalized geometry}\lab{ssshadow}
There are a number of strange things about our construction, one of them being the necessity of introducing wrong-sign kinetic terms for the shadow fields. In this section we will derive it from a gauged linear \si-model that does {\em not} involve negative signature directions -- rather, it replaces the shadow chiral superfields with {\em twisted} chiral superfields. We will also explain the modified Calabi-Yau condition~(\ref{etaCYcond}). This requires a different kind of quotient using the Large Vector Mulitplet introduced in \cite{LVM} which is needed when isometries act {\em simultaneously} on chiral and twisted chiral multiplets\footnote{A different way of gauging such isometries has been proposed in the recent paper~\cite{BKK2025}. The relation between the two is elaborated in~\cite{BLR}.}.
\subsection{The ungauged Linear \si-model}
The generalized K\"ahler potential has the form (\ref{frees}), {\em except} that the negative terms are now interpreted as twisted chiral multiplets $y$:
\beq\lab{freets}
\LL_0=\sum_{i=1}^K|x^i|^2-\sum_{i=K+1}^{K+M}|y ^i|^2~~~,~~~~
\cbar \bbd_+y =\bbd_-y =0~~.
\eeq
As explained in Appendix \ref{geosuprev}, this actually gives the usual Euclidean metric on $\bbc^{K+M}$. The generalized K\"ahler potential has more ambiguities than the usual K\"ahler potential \cite{GHR}: the metric and the closed 3-form are preserved not only under shifts by (the real part of) holomorphic functions $f(x,y )$, but also by (the real part of) twisted holomorphic functions $g(x,\cbar y )$. 

We now focus on one specific shadow field $y $ with charge $q$. Inductively, all shadow fields can be treated in the same way, and so for now we ignore them. We indicate the rest of the Lagrangian by $|\bfx|^2$, where $\bfx$ represents all the remaining chiral fields. We will gauge the following Lagrangian:
\beq\lab{ungauged_twisted}
\LL=|\bfx|^2 -y \cbar y -i\ln\!\left(\frac{\cbar x y }{x\cbar y }\right)\ln\!\left(\frac{x\cbar x}{y \cbar y }\right)~~,
\eeq
where $x$ is an additional chiral field that transforms with the same charge $q$ as $y $. Note that the 
term that we have added is indeed only a total derivative: here $f(x,y )=i\left[\ln\! \left(
\frac{\raisebox{3pt}{$\scriptstyle x$}}{y}\right)\right]^2$ %and $g(x,\cbar y )=2i\ln(x)\ln(\cbar y )$, 
so $x$ appears only as a total derivative and is not a physical field.

\subsection{Gauging with the Large Vector Multiplet.}
The Large Vector Multiplet \cite{LVM,Lindstrom:2008hx} (see Appendix \ref{geosuprev} for a review) is used to gauge isometries that act simultaneously on chiral and twisted chiral superfields. Basically, the idea is to introduce a gauge superfield $V^c$ that transforms with a chiral parameter $\La_c$, a twisted gauge superfield $V^t$ that transforms with a twisted chiral parameter $\La_t$ and a third gauge superfield $V'$ that relates these:\footnote{In \cite{Lindstrom:2008hx}, the notation is slightly different: $V^\phi\lra V^c$, $V^\chi\lra V^t$, $\La\lra\La_c$, and $\tilde\La\lra\La_t$.}
\beq
\de V^c=i(\cbar\La_c-\La_c)~~,~~\de V^t=i(\cbar\La_t-\La_t)~~,~~
\de V' = -\La_c -\cbar\La_c +\La_t+\cbar\La_t~~.
\eeq
Then the gauged \si-model  for (\ref{ungauged_twisted}) is
\bea\label{gauged_twisted}
\LL_V&=& \bar{\bfx}\,e^{Q\cdot V^c}\!\!\cdot\!\bfx-|y|^2e^{qV^t} -\frac{c}2(V^c+V^t)~
\non[2mm]
&&\qquad\qquad-\left[qV'+i\ln\!\left(\frac{\cbar x y }{x\cbar y }\right)\right]
\ln\!\left(\frac{|x|^2 e^{qV^c}}{|y |^2 e^{qV^t}}\right)~.
\eea
Here the first term is meant to represent the gaugings that we have considered before with the charge matrix $Q$ for the particular symmetry that acts on the shadow field $y $; the third term is an FI term. One might expect three such terms, one each for $V^c,V^t,V'$, but the FI term for $V'$ can be removed by a real rescaling of $x$, and similarly, an FI term for $V^c-V^t$ can be absorbed into the phase of $x$, so only one FI term remains.

To recover our previous construction, we simply integrate out $V'$; this replaces $|y |^2 e^{qV^t}$ with $|x|^2 e^{qV^c}$.  Now $x$ has a wrong-sign kinetic term and has become the shadow field.

Now the usual calculation \cite{Witten} that gives the Calabi-Yau condition works. The chiral fields contribute to the FI-term tadpole for $V^c$ with their charge $q$, and the twisted chiral fields contribute to the tadpole for $V^t$ with $-q$, but $V^c-V^t$ is precisely the FI term that is removable (as we noted above).  Details of the superfield Feynman rules are given in Appendix \ref{geosuprev}.

\section{Conclusions}
We extend the Gauged Linear Sigma Model (GLSM) formalism, which in general does not yield Ricci-flat metrics even when the resulting quotient manifold is Calabi–Yau. The essential new ingredient is the introduction of {\em shadow fields}: chiral superfields with kinetic terms of opposite sign. This modification allows us to construct explicit, nonsingular Ricci-flat metrics on the total space of a canonical line bundle over a product of complex projective spaces. A further crucial aspect of our construction is that it can be embedded into a model in which the shadow fields are replaced by physical twisted chiral superfields. This embedding enables a natural generalization of the Calabi–Yau condition, in which the charges are required to sum to zero with an additional relative sign.

\section*{Acknowledgements}
MR is happy to thank P.M.~Crichigno for seminal conversations on this topic many years ago, C.~Vafa for conversations and encouragement, and many other colleagues including S.~Donaldson, C.~Lebrun, Heather Lee, and Yang Li for helpful conversations, as well as the 2021 and 2022 Simons Summer Workshop for providing a stimulating environment. OH would like to thank  Cumrun Vafa for discussions about mirror symmetry and Dario Martelli for pointing us towards many relevant resources. OH, MR, and RvU thank the 2023, 2024, and 2025 Simons Summer Workshops for the chance to work together on this problem. OH and MR also thank the 2026 Simons Summer Workshop. MR thanks NSF grants PHY-1620628 and PHY-2210533 for partial support. The work of RvU and OH is supported by the Czech Science Foundation GACR through the grant “On and Off-shell Superspace Representations” (GA26-23375S). The work of DB was supported by the Russian Science Foundation grant \href{https://rscf.ru/project/7itJjWqd9pYqz47liTvCgk4qUmdtral1n3ecf41C_0I7jWkQGL8qIR8wWeq93dHAgwWeRIZfZPc~/}{№ 25-72-10177}. UL gratefully acknowledges the Simons Summer Workshop 2024, a visiting fellowship at the Simons Center in 2025 along with a visiting professorship at Imperial College sponsored by the Leverhulme trust, and the hospitality of the theory group at Imperial College, as well as support from ``Stiftelsen Lars Hiertas Minne".

\appendix

\section{The Monge-Amp\`ere equation}\lab{MAapp}

Here we write out the Monge-Amp\`ere equation in the special case when the potential is given by the generalized Calabi's ansatz over $\mathcal{B}_{\cp}$, see~(\ref{F_ansatz}).

First, we need to choose $D$ independent coordinates; one choice is
$\mu_i^A,~~i=1\dots(\nsa-1)$, $A=1 \dots N$ and $\mu$, where
\bea
L_i^A&\equiv&\mu_i^A+s^A\qquad\qquad\qquad \hbox{for}~~  i < \nsa~, \non
L^A_{\nsa}&\equiv& s^A-\sum\limits_i^{\nsa-1}\mu^A_i~~,\non
&\hbox{with}& s^A\equiv\mu+a_A~~.
\eea

We calculate the first derivatives of $F_G$ given by~(\ref{F_ansatz}):
\bea
&&\frac{\pa F_G}{\pa \mu^A_i}=-\ln(L^A_i)+\ln(L^A_{\nsa})~~,\non[3mm]
&&\frac{\pa F_G}{\pa \mu}=-\ln(R)
-\sum_A^N\sum_{i=1}^{\nsa}\Big(\ln(L^A_i)-\ln(s^A)\Big)~~.
\eea
In our case, the coefficients $A_\al=0$ except for $A_\mu=1$ in (\ref{MAsymp})
and hence the right-hand side of the Monge-Amp\`ere equation is
\beq
e^{\frac{\pa F_G}{\pa \mu}}=\frac1R\prod_A^N\left(\prod_{i=1}^{\nsa}\frac{s^A}{L^A_i}\right)~.
\eeq
To evaluate the left-hand side of the Monge-Amp\`ere equation (\ref{MAsymp}), we need to calculate the Hessian:
\begin{align}\lab{fij}
F^A_{ij}&\equiv\frac{\pa^2F_G}{\pa\mu^A_i\mu^B_j} =
-\de_{AB} \left(\de^{ij}\frac1{L^A_i}+\frac1{L^A_{\nsa}}\right)
~~,~~\non[3mm]
F^A_{i\mu}&\equiv\frac{\pa^2F_G}{\pa\mu^A_i\pa\mu}=-\frac1{L^A_i}+\frac1{L^A_{\nsa}}
~~,~~\non[3mm]
F_{\mu\mu}&\equiv\frac{\pa^2F_G}{\pa\mu^2}=-\frac{R'}R
-\sum_A^N \sum_{i=1}^{\nsa}\left(\frac1{L^A_i}-\frac1{s^A}\right)~~.
\end{align}
The determinant of the Hessian can be written as
\bea
\det(-\pa\pa F_G)=-\left(\prod_A^N\det(-F^A_{ij})\right)
\left(F_{\mu\mu}-\sum_A^N\sum_{i,j}^{\nsa-1}F^A_{\mu i}(F^A_{ji})^{-1}F^A_{j\mu}\right)~.
%\non[3mm]
\eea
The first factor is (see Appendix~\ref{identities} %(\ref{detM}) 
for details)
\beq
\prod_A^N\det(-F^A_{ij})=\prod_A^N\left[\left(\prod_i^{\nsa}
\frac1{L^A_i}\right)\nsa s^A\right]
\eeq
The second factor is evaluated using $\sum\limits_1^{\nsa-1}L^A_i=\nsa s^A-L^A_\nsa$ and (\ref{invM})
\begin{align}
-\Bigg(F_{\mu\mu}\,-&\sum_A^N\sum_{i,j}^{\nsa-1}F^A_{\mu i}(F^A_{ji})^{-1}F^A_{j\mu}\Bigg)\hspace{5cm}\non[1mm]
=&~\frac{R'}R
+\sum_A^N \sum_{i=1}^{\nsa}\left(\frac1{L^A_i}-\frac1{s^A}\right)\non[1mm]
&~-\sum_A^N\sum_{i,j}^{\nsa-1}\!\Bigg(\frac{-1}{L^A_i} +\frac1{L^A_{\nsa}}\Bigg)\!\Bigg[\!\de^{ij}L^A_i -\frac{L^A_iL^A_j}{\nsa s^A}\Bigg]\!\Bigg(\frac{-1}{L^A_j} +\frac1{L^A_{\nsa}}\Bigg)
\non[1mm]
=&~\frac{R'}R
+\sum_A^N \sum_{i=1}^{\nsa}\left(\frac1{L^A_i}-\frac1{s^A}\right)\non[1mm]
&\qquad-\sum_A^N\sum_j^{\nsa-1}\!\Bigg(\!-1+\frac{L^A_j}{s^A} \Bigg)\!\Bigg(\frac{-1}{L^A_j} +\frac1{L^A_{\nsa}}\Bigg)
\non[1mm] 
=&~\frac{R'}R~~.
\end{align}
Combining all the factors, the Monge-Amp\`ere equation gives the simple result~(\ref{diffR}).

\section{Matrix identities}\lab{identities}
Consider a matrix $M$ which is the sum of a diagonal matrix $D$ and a particular rank 1 matrix~$X$:
\beq
M=D+X~~~\hbox{where}~~~X=x
\left(\begin{array}{c}
1\\
\vdots\\
1
\end{array}\right)\otimes 
\Big(1~\cdots~1\Big)~~.
\eeq
Note that for any diagonal matrix $D$, $\tr(DX)=x\tr(D)$.

We are interested in the determinant and the inverse of $M$.
We start by writing $M=D(1+D^{-1}X)$. The determinant
is
\beq
\det(M)=\det(D)\det(1+D^{-1}X)=\det(D)(1+\tr(D^{-1}X))=
\det(D)(1+x\tr(D^{-1}))~,
\eeq
where for any rank 1 matrix $Y$, we have $\det(1+Y)=1+\tr(Y)$. Similarly, for the inverse we have
\begin{align}
M^{-1}&=(D(1+D^{-1}X))^{-1}~~=~~(1+D^{-1}X)^{-1}D^{-1}\non[1mm]
&=\left(1-\frac{D^{-1}X}{1+\tr(D^{-1}X)}\right)D^{-1}
=\left(1-\frac{D^{-1}X}{1+x\tr(D^{-1})}\right)D^{-1}~~,
\end{align}
where for any rank 1 matrix $Y$, $Y^2=\tr(Y)Y$.
In our case, the matrix that we are interested in has block diagonal form given by (\ref{fij})
\beq
M^A_{ij}=-F^A_{ij}\equiv \de^{ij}\frac1{L^A_i}+\frac1{L^A_{\nsa}}~~
\then~~D^A_{ij}=\de^{ij}\frac1{L^A_i}~~,~~x^A=\frac1{L^A_{\nsa}}~~.
\eeq
Then for each block (dropping the $A$ superscript on $M$), we have
\beq\lab{detM}
\det{M}=\Big(\prod_{i=1}^{\nsa-1}\frac1{L^A_i}\Big)\Big(1+
\frac{\nsa s^A -L^A_{\nsa}}{L^A_{\nsa}}\Big)
=\nsa s^A\prod_{i=1}^{\nsa} \frac1{L^A_i}~~,
\eeq
and
\beq\lab{invM}
M^{-1}_{ij}=\de^{ij}L^A_j -\frac{L^A_iL^A_j}{\nsa s^A}
\eeq
\section{Geometry and Superspace}\lab{geosuprev}
The relation between complex geometry and superspace is well known, e.g. \cite{Zumino,Howe,HKLR}, but we briefly review it here. We work in $(2,2)$ superspace characterized by complex left and right spinor derivatives obeying the superalgebra:
\beq
\bbd_\pm^2=\cbar \bbd_\pm^2=0~~,~~~ \{\bbd_\pm,\cbar \bbd_\pm\}=i\pa_{\pm\pm}
~~,~~~ \{\bbd_\pm,\cbar \bbd_\mp\}=0~~.
\eeq
\subsection{Superfields}
\subsubsection{Matter fields}
Superfields that we consider are of several types: chiral superfields $x^i,z^\al\dots$,  twisted chiral superfields $y ^i$, and real gauge superfields $V_a$.
The chiral and twisted chiral superfields obey
\beq\label{xy}
\cbar \bbd_\pm x=\cbar \bbd_\pm z=0~~,~~~ \cbar \bbd_+y =\bbd_-y =0~~,
\eeq
and their complex conjugates.

Other superfield constraints are interesting \cite{GHR}; in particular, we note that $\mu=y +\cbar y $
obeys
\beq
\cbar \bbd_+ \cbar \bbd_-\mu = \bbd_+ \bbd_-\mu=0~~,
\eeq
and hence $\mu$ is called a real linear superfield. The superfields corresponding to $L$ in the text are real linear superfields as well. 

\subsubsection{Standard Vector Multiplet}
The gauge fields are real and unconstrained 
and transform as
\beq\lab{Vchiral}
\de V = i(\cbar\La_c -\La_c)~~,~~~\cbar \bbd_\pm \La_c=0
\eeq
where $\La_c$ is a chiral parameter. The gauge field $V$ can be used to construct gauge-covariant spinor derivatives; in chiral representation, these are:
\begin{align}
&\cbar\bmn_\pm=\bbDB{\pm}~,~~{\bmn}_\pm=e^{-qV}\bbD{\pm}e^{qV}
\end{align}
where $q$ is the charge of the field the derivative acts on.
These obey the algebra
\begin{align}\lab{bmnalg}
&\{\bmn_\pm,\cbar\bmn_\pm\}
=i\bmn_{\pm\pm}~~,~~
\{\cbar\bmn_+,\bmn_-\}=iqF
~~,~~
\{\bmn_+,\cbar\bmn_-\}=iq\bar F~~,\non[2mm]
&\bmn_\pm^2=\cbar\bmn_\pm^2= 
\{\bmn_+,\bmn_-\} =  
\{\cbar\bmn_+,\cbar\bmn_-\} =0~~,
\end{align}
where $F$ is a twisted chiral field strength.

\subsubsection{Large Vector Multiplet}
The Large Vector Multiplet (LVM) 
\cite{LVM,Lindstrom:2008hx} is used to gauge isometries that act simultaneously on chiral and twisted chiral superfields. In the Abelian case, it is described by three real gauge superfields: $V^c$ that transforms with a chiral parameter $\La_c$, $V^t$ that transforms with a twisted chiral parameter $\La_t$ and a $V'$ that relates these (in the conventions of \cite{Lindstrom:2008hx}):
\beq
\de V^c=i(\cbar\La_c-\La_c)~~,~~\de V^t=i(\cbar\La_t-\La_t)~~,~~
\de V' = -\La_c -\cbar\La_c +\La_t+\cbar\La_t~~.
\eeq
As in \cite{Lindstrom:2008hx}, we also define the complex combinations
\begin{align}
&V_L:=\frac 1 2 \big[-V'+i(V^c-V^t)\big]~,~~V_R:=\frac12 \big[-V'+i(V^c+V^t)\big]~,\non
&~~~\,\Rightarrow ~\de V_L=\La_c-\La_t~,~~\de V_R=\La_c-\cbar\La_t~~~,~~~V_L=V_R-iV^t~,~~\bar V_L=V_R-iV^c~~.
\end{align}
In chiral representation, all covariant derivatives transform with the chiral parameter $\La_c$ and we have
\begin{align}
&\cbar\bmn_\pm^c=\bbDB{\pm}~,~~{\bmn}_\pm^c=e^{-qV^c}\bbD{\pm}e^{qV^c}
\end{align}
with $q$ the charge of the superfield being acted on. For chiral fields $x$ and twisted chiral fields $y$ and their complex conjugates, the charges are denoted 
\begin{align}\lab{covcon}
&q_{x}=-q_{\bar x}~,~~q_{y}=-q_{\bar y}~,
\end{align}
respectively.
These derivatives obey the same algebra as above (\ref{bmnalg})\footnote{This definition of $F,\tilde F$ differs from the definition in \cite{Lindstrom:2008hx} (3.16).}.

In chiral representation, the ``twisted'' covariant derivatives are (\cite{Lindstrom:2008hx} (3.8))
\begin{align}\label{td}
&{\bmn}_+^t=e^{iqV_R}\bbD{+}e^{-iqV_R}~,&\bar{\bmn}_+^t=e^{iqV_L}\bbDB{+}e^{-iqV_L}\non[1mm]
&\bar{\bmn}_-^t=e^{iqV_R}\bbDB{-}e^{-iqV_R}~,&{\bmn}_-^t=e^{iqV_L}\bbD{-}e^{-iqV_L}
\end{align}
In terms of the covariant derivatives the spinorial field strengths are given by 
\begin{align}\label{gpm}
&\bar{\bmn}^c_+-\bar{\bmn}^t_+:= iq\bbg_+=iq\bbDB{+}V_L~,~~~\bmn^c_+-\bmn^t_+:= iq\cbar\bbg_+=iq\bbD{+}\bar V_L \non[1mm]
&\bar{\bmn}^c_--\bar{\bmn}^t_-:= iq\bbg_-=iq\bbDB{-}V_R~,~~~\bmn^c_--\bmn^t_-:= iq\cbar\bbg_-=iq\bbD{-}\bar V_R~~.
\end{align}
The algebra of the full set $\bmn^t,\bmn^c$ follows from this and the algebra of $\bmn^c$ above. In particular, from the explicit form (\ref{td}), we have
\begin{align}\label{algt}
&\{\cbar\bmn_+^t\,,\,\bmn_-^t\}=0~~\then~~
\{\cbar\bmn_+^c\,,\,\bmn_-^c\}-iq(\bbDB+\cbar\bbg_-+\bbd_-\bbg_+)=0\non[1mm]
&\then~~ F=\bbDB+\cbar\bbg_-+\bbd_-\bbg_+
\non[1mm]
&\{\cbar\bmn_+^t\,,\,\cbar\bmn_-^t\}=iq\tilde F ~~\then~~ \tilde F=-(\bbDB+\bbg_-+\bbDB-\bbg_+)
~~,~~etc.
\end{align}
\subsubsection{Charged Matter Fields}
Chiral fields satisfy $\bbDB{\pm}x=\bbD{\pm}\bar x=0$, while twisted chiral fields satisfy
$\bbDB{+}y=\bbD{-}y=0$ and $\bbD{+}\bar y=\bbDB{-}\bar y=0$. They are charged under $\La_c$ and $\La_t$ transformations. In chiral representation the fields should all transform with $\La_c$. From the definitions of $V_L$ and $V_R$ we deduce that the covariant versions are
\begin{align}\label{creps}
&\hat x= x
~,~~\hat {\bar x}
=\bar x e^{-q_{\bar x} V^c}~,~~
\hat y=y e^{iq_y V_L}~,~~{\hat {\bar y}}=\bar y e^{iq_{\bar y}V_R}~~.
\end{align}
These covariantly (twisted) chiral fields then transform as
\begin{align}
&\hat x'=e^{iq_x\La_c}\hat x~,~~~\hat{\bar x}'=e^{iq_{\bar x}\La_c}\hat{\bar x}~,~~\hat y'=e^{iq_y\La_c}\hat y~,~~~\hat{\bar y}'=e^{iq_{\bar y}\La_c}\hat{\bar y}
\end{align}
and satisfy the covariant versions of the chirality constraints
\begin{align}\label{covcons}
\bar{\bmn}^c_{\pm}\hat x=\bmn^c_{\pm}\hat{\bar x}=
\bar{\bmn}^t_{+}\hat y=\bmn^t_{-}\hat y=
{\bmn}^t_{+}\hat{\bar y}=\bar{\bmn}^t_{-}\hat{\bar y}
=0~~.
\end{align}
Note that nothing dictates a relation between the charges of the chiral and twisted chiral superfields. In section \ref{ssshadow}, we choose $q_x=q_y$ and hence $x\bar y$ and $x/y$ are gauge-invariant combinations. 

In terms of the chiral representation superfields (\ref{creps}), the gauged Lagrangian (\ref{gauged_twisted}) has the form of the ungauged theory (\ref{ungauged_twisted}) with all the superfields in the chiral representation:
\begin{align}\lab{gauged_twisted_chiral_rep}
\LL_V &= |\hat\bfx|^2 -\hat y\hat{\bar y} -i\ln\!\left(\frac{\hat{\bar x} \hat y }{\hat x\hat{\bar y} }\right)\ln\!\left(\frac{\hat x\hat{\bar x}}{\hat y \hat{\bar y} }\right)+\frac{c}{2q_x}\ln(\hat x\hat{\bar x}\hat y\hat{\bar y})\non[2mm]
&=|\hat\bfx|^2 -\hat y\hat{\bar y}+i\left[\ln\!\left(\frac{\hat x}{\hat y}\right)\right]^2-i\left[\ln\!\left(
\frac{\hat{\bar x}}{\hat{\bar y}}\right)\right]^2
+\frac{c}{2q_x}\ln(\hat x\hat{\bar x}\hat y\hat{\bar y})
\end{align}
This form makes the $(1,1)$ superspace decomposition straightforward (see below).

\subsection{Supergraphs}
To compute anomalies, we need the Feynman rules for chiral and twisted chiral superfields. The propagators in $(2,2)$ superspace follow from the Lagrangian (\ref{freets}). As was shown in \cite{Supergraphs}, chiral multiplets have a propagator 
\beq
\underbracket{x(1)\,\cbar x(2)}=\frac{\de^4(\te_1-\te_2)}{k^2+m^2}~~,
\eeq
with a factor $\cbar \bbd_+\cbar \bbd_-$ for each $x$ in any vertex and a factor $\bbd_+\bbd_-$ for each $\cbar x$ in any vertex; here $m$ acts as an infrared regulator. Thus at a
vertex $|x|^2Q\cdot V$ which comes from expanding $|x|^2e^{Q\cdot V}$, we would have 
\beq
Q\cdot V(1,k)\frac{\cbar \bbd_+\cbar \bbd_-\bbd_+\bbd_-\,\de^4(\te_1-\te_2)}{k^2+m^2}
~
\eeq
modulo terms with fewer internal spinor derivatives, i.e., with some acting on $V$.

Similarly, twisted chiral multiplets have a propagator (taking into account the $-$ sign in the Lagrangian)\cite{Buscher85}
\beq
\underbracket{y (1)\,\cbar y (2)}=-\frac{\de^4(\te_1-\te_2)}{k^2+m^2}~~,
\eeq
with a factor $\cbar \bbd_+\bbd_-$ for each $y $ in any vertex and a factor $\bbd_+\cbar \bbd_-$ for each $\cbar y $ in any vertex. Thus at
vertex $-|y |^2qV$ which comes from expanding 
$-|y |^2e^{qV}$, we would have 
\bea
&[-qV(1,k)]\left[-\frac{\cbar \bbd_+\bbd_-\bbd_+\cbar \bbd_-\,\de^4(\te_1-\te_2)}{k^2+m^2}\right]\qquad\qquad&
\non[2mm]
&\qquad\qquad~\sim~ -qV(1,k)\frac{\cbar \bbd_+\cbar \bbd_-\bbd_+ \bbd_-\,\de^4(\te_1-\te_2)}{k^2+m^2}~,&
\eea
where the final $-$ sign comes from reordering the spinor derivatives and we again drop terms with fewer spinor derivatives. Thus we see that {\em twisted chiral multiplets contribute to the tadpole calculation with a sign opposite to the contribution of chiral multiplets}; this sign difference was already observed in \cite{Buscher85}.

\subsection{(1,1) components}\lab{components}
Actions in (1,1) superspace can be understood much more directly, as all the superfields are unconstrained, and metrics, moment maps, etc., all appear explicitly. Here we consider the reduction from (2,2) to (1,1) superspace (largely) using the notation of \cite{LVM}. We start by defining 
\beq
\bbd_\pm=\frac12(D_\pm-iQ_\pm)~~,~~ \bbbd_\pm=\frac12(D_\pm+iQ_\pm)~~,
\eeq
where $D_\pm$ are the (1,1) spinor derivatives and $Q_\pm$ are the generators of the extra supersymmetries ({\em not} the (1,1) supercharges). They obey
\beq
D_\pm^2=Q_\pm^2=i\pa_{\pm\pm}~~,
\eeq 
with all other anticommutators of $D,Q$ vanishing.
\subsubsection{Chiral and twisted chiral fields}
The (1,1) components of chiral and twisted chiral fields are simply complex (1,1) scalar superfields. The role of their (2,2) constraints is to prescribe their transformations under the extra supersymmetries generated by $Q_\pm$. Consider a chiral superfield $x$ and a twisted chiral superfield~$y$:
\bea
&&\bbbd_\pm x=0 \then Q_\pm x = iD_\pm x~~,~~
\bbd_\pm \bar x=0 \then Q_\pm \bar x = -iD_\pm \bar x~~,\non
&&\bbbd_+y=\bbd_-y=0 \then Q_+y=iD_+y~~,~~Q_-y=-iD_-y~~,~~etc.
\eea
Using $D_\pm\equiv\bbd_\pm+\bbbd_\pm$ and the chirality constraints on $x,y$, we can rewrite these as
\beq\lab{chit11}
\bbd_\pm x= D_\pm x~~,~~
\bbbd_\pm \bar x = D_\pm \bar x~~,~~
\bbd_+ y =D_+ y~~,~~
\bbbd_-y=D_-y~~,~~
\bbbd_+\bar y=D_+\bar y~~,~~
\bbd_-\bar y =D_-\bar y~~.
\eeq
When we gauge symmetries that act on these fields, it is much more useful to define gauge-covariant (1,1) components using (\ref{covcons}). 

\subsubsection{Standard Vector multiplet}
To find the (1,1) superspace components of the vector multiplet, we have two choices--we can find a Wess-Zumino gauge or proceed covariantly. Here we do the latter. We define the (1,1) covariant derivative as
\beq\label{vec11}
\na\!_{\pm}:=\bmn_\pm +\cbar\bmn_\pm~. 
\eeq
The covariant (1,1) spinor derivative algebra is defined in the usual way:
\beq\label{vec11alg}
\na\!_{\pm}^{\,\,2}=i\,\na\!_{\pm\pm}~~,~~~
\{\na\!_{+},\na\!_-\}= iqf~~,
\eeq
where the (1,1) field strength $f$ is related to the (2,2) field strengths (\ref{bmnalg})
\beq
f=F+\bar F~~;
\eeq
the (1,1) description also contains an independent unconstrained (1,1) scalar $s$ which descends from the (2,2) field strengths:
\beq
s=i(\bar F -F)
\eeq
\subsubsection{Large Vector multiplet}
To reduce the LVM to (1,1) superspace,
we start by choosing which of the many covariant spinor derivatives of the LVM will be used for the (1,1) derivatives. Because we have (2,2) gauge-invariant spinor field strengths (\ref{gpm}), the definition of the gauge connections is ambiguous; we choose to simplify the part of the Lagrangian with chiral fields, which is a different choice than in \cite{Lindstrom:2008hx}\footnote{The choices here are convenient only for the Abelian case.}, and simply follow the conventions of the standard vector multiplet above (\ref{vec11}):
\beq\label{def11}
\na\!_{\pm}:=\bmn_\pm^c +\cbar\bmn_\pm^c~.
\eeq
The covariant (1,1) spinor derivative algebra is again defined in the usual way:
\beq\label{fdef}
\na\!_{\pm}^{\,\,2}=i\,\na\!_{\pm\pm}~~,~~~
\{\na\!_{+},\na\!_-\}= iqf~~,
\eeq
and the relation to the (2,2) field-strengths follows immediately from the definition (\ref{def11}). 

There are some useful tricks for reducing the (2,2) action to (1,1) superspace. Using
\beq
Q_\pm=i(2\bbd_\pm -D_\pm)=-i(2\cbar\bbd_\pm -D_\pm)~~,~~
\bbd_\pm=-\cbar\bbd_\pm+D_\pm~~,
\eeq
as well as $D_\pm^2=i\pa_{\pm\pm}$, up to total derivatives we may write the (2,2) measure as
\begin{align}\label{mess}
&\int\cbar\bbd_+\cbar\bbd_-\bbd_+\bbd_- =-\int\cbar\bbd_+\bbd_-\bbd_+\cbar\bbd_-=\int D_+D_-\bbd_+\bbd_-=\int D_+D_-\cbar\bbd_+\cbar\bbd_-~\nn[2mm]
&=-\int D_+D_-\bbd_+\cbar\bbd_-=-\int D_+D_-\cbar\bbd_+\bbd_-=-{\textstyle\frac14}\int D_+D_-Q_+Q_-~~.
\end{align}
The different forms of the measure streamlines finding the (1,1) components because of the chirality properties of the superfields.

As usual, $\bbd_\pm$ are covariant when acting on gauge-invariant expressions, and hence using Leibniz rule, on superfields, they turn to $\bmn^c_\pm$. Then we use (\ref{covcons}), (\ref{gpm}), (\ref{def11}), and $q=q_x=q_y=-q_{\bar x}=-q_{\bar y}$ to find the covariant version of (\ref{chit11}):
\begin{align}\label{covrels}
&\bmn^c_\pm\hat{\bar x}=0~~,\qquad\qquad
\bmn^c_\pm\hat{x}=\na\!_{\pm} x~~,~\qquad\qquad\cbar\bmn^c_\pm\hat{\bar x}=\na\!_{\pm}{\bar x} ~~,\qquad\,
\cbar\bmn^c_\pm\hat{x}=0~~,\non[2mm]
&
\bmn^c_+\hat{\bar y}=-iq\cbar\bbg_+{\bar y}~~,~~~
\bmn^c_+\hat{y}=\na\!_{+} y -iq\bbg_+ y ~~,~~
\cbar\bmn^c_+\hat{y}=iq\bbg_+{y}~~,
\,\,~~~
\cbar\bmn^c_+\hat{\bar y}=\na\!_{+}{\bar y}+iq\cbar\bbg_+{\bar y}~~,
\non[2mm]
&
\bmn^c_-\hat{\bar y}=\na\!_-{\bar y}+iq\bbg_-{\bar y}
~~,~~~\bmn^c_-\hat{y}=iq\cbar\bbg_-{y}
~~,~~~
\cbar\bmn^c_-\hat{\bar y}=-iq\bbg_-{\bar y}~~,~~~
\cbar\bmn^c_-\hat{y}=\na\!_- y -iq\cbar\bbg_-y~~,
\end{align}
where the (1,1) components are defined in terms of the covariant (2,2) superfields, so, e.g., $\hat x \to x$, etc.

Recall that in the Abelian case, field strengths are gauge-invariant, so $\bmn \bbg=\bbd\bbg$, etc. Then 
the (1,1) components of the LVM are given by 
\beq
\bbg_\pm~~,~~\cbar\bbg_\pm~~,~~
S:=\bbd_-\bbg_+~~,~~
\bar{\tilde S} :=\bbd_+\bbg_-~~,~~
\bar S:=\bbbd_-\cbar\bbg_+~~,~~
\tilde S:=\bbbd_+\cbar\bbg_-~~,
\eeq
and (1,1) derivatives of these.\footnote{For completeness, we note
$\cbar\bbd_\pm \bbg_\pm=\bbd_\pm\cbar\bbg_\pm=0$ (though it plays no role in going to (1,1) components).}
We use (\ref{def11}) to rewrite, e.g.,
$\bbd_\pm\cbar\bbg_\mp=D_\pm \cbar\bbg_\mp -\bbbd_\pm \cbar\bbg_\mp$.
We also need to recall the definition of the (1,1) field strength $f$ (\ref{fdef}), which gives the relation
\beq\label{fdef11}
f=S +\tilde S+ \bar S+\bar{\tilde S}~~.
\eeq
\subsubsection{The action in (1,1) superspace}
All this is enough to find the (1,1) form of the superspace action (\ref{gauged_twisted_chiral_rep}). 
The various forms of the measure in (\ref{mess}) lets us choose a form adapted to the chirality properties of the term we are evaluating. For the chiral superfield term $|\hat\bfx|^2$, we use:
\beq\label{xterm}
-\,\half\int(\bbbd_+\bbd_-+\bbd_+\bbbd_-)|\hat\bfx|^2=-\,\half\int (\na\!_{+}\bar \bfx
\cdot \na\!_- \bfx+\na\!_{+}\bfx
\cdot \na\!_- \bar \bfx)-\bar \bfx\cdot (Qs)\,\bfx~~,
\eeq
where $Q$ is the charge matrix acting on the chiral fields (\ref{gauged_twisted}), $\int$ is the (1,1) measure $\int d^2\sigma \, D_+D_-$,  and 
\beq\label{sdef}
s=i(\cbar F-F)=i(\bar S+\bar{\tilde S}-S-\tilde S)~~.
\eeq

Next, we consider the twisted superfield term $-|\hat y|^2$:
\begin{align}\label{yterm}
&\half\int(\bbbd_+\bbbd_-+\bbd_+\bbd_-)(-|\hat y|^2)\non[1mm]
&\qquad=-\,\half\int \Big((\na\!_{+}+iq\Xi_+)\bar  y\,
(\na\!_--iq\Xi_-)  y+(\na\!_{+}\! -iq\Xi_+)y\,
( \na\!_-+ iq\Xi_-)\bar  y- q\tilde s|y|^2\Big)~~,
\end{align}
where
\begin{align}
\tilde s:=i(\cbar{\tilde F}-\tilde F)&=i(\bbDB+\bbg_-+\bbDB-\bbg_+-\bbD+\cbar\bbg_--\bbD-\cbar\bbg_+)\non[1mm]
&=-(D_+\tilde\Xi_-+D_-\tilde\Xi_+)+i(\bar S+\tilde S -S-\bar{\tilde S})~~,
\end{align}
and $\Xi_\pm:=\bbg_\pm+\cbar\bbg_\pm$,
$\tilde\Xi_\pm:=i(\cbar\bbg_\pm-\bbg_\pm)$.

The FI-term gives
\beq\label{FIterm}
\frac{c}{2q}\left[-\half\int(\bbbd_+\bbd_-+\bbd_+\bbbd_-)\ln(|\hat x|^2) +
\half\int(\bbbd_+\bbbd_-+\bbd_+\bbd_-)\ln(|\hat y|^2)\right]=\frac{c}4\int(s-\tilde s)~~.
\eeq

Finally, we consider the $[\ln(\frac{\hat x}{\hat y})]^2$ term:
\begin{align}\label{novterm}
   -i \int \bbbd_+\bbd_- \left[\ln\!\left(\frac{\hat x}{\hat y}\right)\right]^2 &=-2iq\int \bbd_-\left[\ln\!\left(\frac{\hat x}{\hat y}\right)(i\bbg_+)\right]\non[2mm]
  &=2q\int \left[\ln\!\left(\frac{x}{y}\right)\!S
  +[\na\!_-\ln(x)]\bbg_++iq\,\bbg_+\cbar\bbg_-\right]~~,
\end{align}
and its complex conjugate:
\begin{align}
   &i \int \bbd_+\bbbd_- \left[\ln\!\left(\frac{\hat{\bar{x}}}{\hat {\bar{y}}}\right)\right]^2 =2q\int \left[\ln\!\left(\frac{\bar x}{\bar y}\right)\!
   \bar S
  +[\na\!_-\ln(\bar x)]\cbar\bbg_+-iq\,\cbar\bbg_+\bbg_-\right]~~.
\end{align}
\subsubsection{Eliminating redundant fields}
We want to see how the shadow fields arise in (1,1) superspace. To do this, we need to eliminate all the extra fields that come from the LVM and leave only the (1,1) components $s,f$ and $\na\!_\pm$ of the usual (2,2) vector multiplet. This means that we integrate out all the $\bbg_\pm,\cbar\bbg_\pm$, and, due to (\ref{fdef11},\ref{sdef}), two of $S,\tilde S,\bar S,\bar{\tilde S}$. We choose $S,\bar S$ and write
\beq
\tilde s = -s + 2i(\bar S-S) - 
(D_+\tilde \Xi_-+D_-\tilde\Xi_+)~~.
\eeq

Integrating out $S,\bar S$ gives:
\beq
-iq|y|^2+\frac{ic}2+2q\ln\!\left(\frac{x}{y}\right)=0~~,~~
iq|y|^2-\frac{ic}2+2q\ln\!\left(\frac{\bar x}{\bar y}\right)=0~~,
\eeq 
which implies
\beq
|y|^2=|x|^2~~,~~\frac{y}{\bar y}=
e^{i\left(\frac{c}{2q}-|y|^2\right)}\frac{x}{\bar x}
~~\then~~y=e^{\ihalf\left(\frac{c}{2q}-|x|^2\right)}x:=e^{i\varphi}x~~.
\eeq
Thus $x$ is $y$ up to a phase, and has the same charge $q$ (which is consistent with the choice we made in (2,2) superspace in section \ref{ssshadow}).

Integrating out $\bbg_+,\cbar\bbg_+$ gives
\beq\label{}
q\big(2i\cbar\bbg_--\Xi_-|y|^2\big)=2\na\!_-\ln x-iy\na\!_-\bar y~,
\eeq
and its complex conjugate. 
We may rewrite these in terms of $x$ using
\beq
y\na\!_-\bar y={\textstyle\frac i 4} D_-|x|^4+x\na\!_-\bar x~.
\eeq
It follows that
\beq
q\Xi_-=-i\na\!_-\ln \frac {x}{\bar x}-\half D_-|x|^2
\eeq
and 
\beq
iq\cbar\bbg_-=\na\!_-\ln x -\ihalf\bar x \na\!_-x~~\then~~q\tilde\Xi_-=D_-\ln|x|^2+\ihalf
(x\na\!_-\bar x -\bar x \na\!_- x)~~.
\eeq
Because $\bbg_+,\cbar\bbg_+$ (as well as $S,\bar S$) enter the action at most linearly, they act as Lagrange multipliers, and we do not need their expressions.

We now substitute these results into (\ref{yterm},\ref{FIterm})\footnote{The remaining terms in the (1,1) reduction of the action vanish because they are proportional to the Lagrange multipliers.} and find the final (1,1) action that results from our novel LVM quotient. We use the following intermediate results:
\beq
\na\!_\pm y =e^{i\varphi}(\na\!_\pm x -\ihalf x D_\pm|x|^2)~~\then ~~
(\na\!_--iq\Xi_-)y=e^{i\varphi}\frac{x}{\bar x}\na\!_-\bar x~~,
\eeq
as well as their complex conjugates. Then 
\beq
-\half\na\!_+\bar y (\na\!_--iq\Xi_-)y=
-\half(\na\!_+\bar x+\ihalf \bar x D_+|x|^2)\frac{x}{\bar x}\na\!_-\bar x~~, 
\eeq
its complex conjugate, and term that comes from $\tilde s$
\beq
-\half q( s |x|^2 -(D_+|x|^2)\tilde\Xi_-)
\eeq
combine to give precisely the shadow action:
\beq
+\half\int (\na\!_{+}\bar x
\na\!_- x+\na\!_{+}x
\na\!_- \bar x)-qs|x|^2~~.
\eeq
The FI term becomes 
\beq
\frac{c}2\int s~~.
\eeq
Thus we have found that the (1,1) superspace quotient which comes from the (2,2) LVM gauging of a theory with no shadow fields but a novel gauged generalized K\"ahler transformation, after eliminating various (1,1) Lagrange multiplier fields, results in a GLSM with shadow fields.

From the (1,1) point of view, this result is very surprising. The terms (\ref{xterm},\ref{yterm},\ref{FIterm}) in the (1,1) action are all conventional with the usual signs (albeit with a shifted connection in (\ref{yterm})). However, the terms (\ref{novterm}) coming from the novel gauging of the generalized K\"ahler transformation change everything. Not unexpectedly, they identify $y$ with $x$ up to a phase, but they also lead to a change in the sign of the kinetic term. 

\subsection{Geometry}
Superspace Lagrangians depending only on chiral superfields and their complex conjugates are the K\"ahler potentials for K\"ahler manifolds with holomorphic coordinates $x,z...$ \cite{Zumino}; likewise superspace Lagrangians depending only on twisted chiral superfields and their complex conjugates are (minus) the K\"ahler potentials for K\"ahler manifolds with holomorphic coordinates $y$. This is the basis of T-duality and mirror symmetry (cf.~\cite{BuscherA,BuscherB,RVdual,Morrison,Hori}). 

Except when the chiral and the twisted chiral superfields don't interact, such as the case considered in (\ref{freets}), superspace Lagrangians that depend on both chiral and twisted chiral multiplets do not give rise to K\"ahler geometry, but rather to a special case of generalized K\"ahler geometry \cite{GHR,BuscherC}.

Whereas chiral superfields correspond to holomorphic coordinates in K\"ahler geometry, the linear superfields $L$ and $\mu$ are identified with the real symplectic coordinates entering the the symplectic potential discussed in section \ref{sec:sympdual}.

Aside for the discussion in section \ref{ssshadow} and the corresponding parts of this appendix\footnote{The geometry of this novel generalized gauging by the LVM is under investigation.}, the quotients in this paper are all K\"ahler quotients, and corresponding to each $U(1)$ action, there is a gauge superfield $V$.
The quotient is performed by extremizing the gauged action with respect to each $V$. If we solve the resulting equations for the $V$'s, since the gauge parameter $\La_c$ is chiral, the gauge symmetry is complexified--in algebraic geometry, this is called the GIT quotient. The points for which we {\em can} solve for $V(z,\cbar z)$ are called semistable. Alternatively, we can use the imaginary part of $\La_c$ to gauge away parts of the $V$'s and go to Wess-Zumino gauge; in this gauge, the $\te$-independent part of the equations are the moment-map equations and constrain the coordinates $x$; the gauge group is now just the usual product of $U(1)$'s, and it preserves the moment-map constraints. This is the symplectic quotient perspective.

\setstretch{0.8}
\setlength\bibitemsep{5pt}
\printbibliography
\end{document}